\documentclass[aps,prd,onecolumn]{revtex4}

\usepackage{graphicx}  
\usepackage{rotating}

\def\lea{\mathrel{\raise .4ex\hbox{\rlap{$<$}\lower 1.2ex\hbox{$\sim$}}}}
\def\gea{\mathrel{\raise .4ex\hbox{\rlap{$>$}\lower 1.2ex\hbox{$\sim$}}}}

\let\gsim=\gea

\begin{document}

\title{Gaussianity\footnote{Chapter of the book ``Cosmic Microwave Background:
from quantum fluctuations to the present universe'', IAC Winter School 2007.}}

\author{Enrique Mart\'\i nez-Gonz\'alez}
\affiliation{Instituto de F\'\i sica de Cantabria (CSIC-UC), Av. Los
  Castros s/n, 39005 Santander, Spain\\
email: martinez@ifca.unican.es}
\date{17th April 2008}


\begin{abstract}
In this chapter we review the present status of the Gaussianity
studies of the 
CMB anisotropies, including physical effects producing
non-Gaussianity, methods to test it and observational constraints. 
\end{abstract}

\maketitle
\tableofcontents

\section{Introduction \label{sec:introduction}}

The study of the CMB temperature and polarization anisotropy has had an
essential 
role in establishing the standard cosmological model: a flat
$\Lambda$CDM model composed of baryonic matter ($\approx 4\%$), dark matter
($\approx 24\%$) and dark energy ($\approx 72\%$), with nearly scale-invariant
adiabatic fluctuations in the energy-density. The information
extracted from the CMB to derive the cosmological model has been
mostly based on the CMB anisotropy power spectrum. However, much more
information is still waiting to be extracted from the statistics of
the CMB beyond the second-order moment.  

The study of the Gaussianity of the CMB anisotropies has become in 
recent years a very relevant topic in the CMB field. The main reason  
for that is the availability of high sensitivity and high resolution maps
provided by the new 
generations of CMB experiments, with which the Gaussian
character of the anisotropies predicted by the standard inflationary
paradigm can be tested at a high accuracy. CMB
anisotropies form an ideal data set for testing the Gaussianity of the
primordial energy-density fluctuations since the
dominant physical effects producing the anisotropies involves just linear
physics, and Gaussianity is preserved under linear
transformations. Another approach to test it is by studying the large
scale structure of the universe as described by the galaxy
distribution. The density perturbations in the linear regime should
also be a good representation of the initial conditions. However,
galaxy formation is a very non-linear process 
involving complex physical effects which very much complicate the
analysis. 

Different models of the early universe have been proposed to naturally
account for the early stages of its history and, thus, not to have to
rely on ad hoc initial conditions. They are based on
different theories, like string theory or M theory, and include or
not an inflationary phase (an example of the latter being the
Ekpyrotic cosmology \cite{khoury_01}). Most of those models predict
very specific 
properties about the probability distribution of the CMB
anisotropies, which in many cases imply deviations from Gaussianity
with amplitudes within reach of present or near future experiments.  
Some of those models are already constrained by present data and many
more are expected to be disproved in the coming years, specially with
the launch of the Planck satellite at the end of 2008 (or beginning of
2009). 

The analysis of the angular distribution of the CMB anisotropies is a
formidable task with profound consequences on our understanding of the
universe. Before any meaninful result can be achieved, it is, however,
crucial to 
control at a high level of precision all possible systematics that can
be introduced by the experiment and the pipeline process used to
reduce the data. On the other hand, there are several Galactic and
extragalactic emissions in the microwave band which blur the
CMB signal. Disentangling the cosmic signal from the others is also
crucial and requires a very good knowledge of the astrophysical
emissions. A lot of observational and theoretical effort has been dedicated    
to that aim in recent years. Below we summarize the present status of
the most important aspects of the analysis of Gaussianity and the
results achieved. 

\section{The isotropic Gaussian random field (IGRF)\label{sec:IGRF}}

One of the most robust predictions of the standard inflationary model
is that the CMB anisotropies should be well represented by an
isotropic Gaussian random field (IGRF) on the celestial sphere. This is a very
powerful prediction to test the standard model. There are many
statistical quantities for a IGRF that can be derived analytically,
which greatly facilitates the Gaussianity test of the CMB
data. However, a disadvantage may come from the central limit
theorem that implies that the sum of several independent non-Gaussian
distributions tends to a more Gaussian one. As we will see below, this
complicates the Gaussianity analysis since the observational data is
composed of several contributions which can be either intrinsic or
extrinsic to the CMB anisotropies.
     
\subsection{Definition \label{subsec:IGRF_definition}}

A random field defined on a given support is said to be Gaussian if
for any N points of the support $x_1, ..., x_n$ the values of the
random field 
$y_1, ..., y_n$ follow a multinormal distribution
\begin{equation}
f(y_1, ...,
y_n)=\frac{1}{(2\pi)^{n/2}|M|^{1/2}} \exp \left(-\frac{1}{2} \sum_{ij}
(y_i-<y_i>)M_{ij}^{-1}(y_j-<y_j>)\right)
\end{equation}
where M is the covariance matrix defined as
$M_{ij}=<(y_i-<y_i>)(y_j-<y_j>)>$. Thus the n-pdf (probability density
function), and also all the moments, is given in terms of just the first two
moments. 

In the case of the temperature anisotropies of the CMB the mean value
is set to zero, and thus the standard model predicts a Gaussian pdf
characterized by only the second
moment. The support is in this case the 2D sphere. Thus the temperature
anisotropies can be expanded in terms of the
spherical harmonic coefficients, 
\begin{equation}
\Delta({\bf n})\equiv \frac{\Delta T}{T}({\bf n})=\sum_{lm} a_{lm}
Y_{lm}(\bf n) 
\end{equation}
where $Y_{lm}(\bf n)$ are the spherical harmonic functions for
direction $\bf n$ and $a_{lm}$ are the spherical harmonic coefficients
which for the standard model are Gaussian distributed .

If the field is isotropic then the two-point correlation function only
depends 
on the modulus of the difference of the two directions. In harmonics
space, isotropy translates in that the harmonic coefficients are
uncorrelated 
\begin{equation}
<a_{lm}a^*_{l'm'}>=C_l \delta_{ll'}\delta_{mm'}
\end{equation}       
where $\delta_{ij}$ is the Kronecker delta. $C_l$ is the temperature
power spectrum which for a realization can be estimated as $(2l+1)^{-1}\sum_l
|a_{lm}|^2$.
 
\subsection{Properties \label{subsec:IGRF_properties}}

The IGRF is one of the best studied random fields and many of its properties
have been thoroughly analysed. One of the most remarkable properties is
that the expectation of any even combination of the field $\Delta(\bf
n)$ can be given in 
terms of the  second moment, the 2-point correlation function, and the
expectation of any odd
combination is zero; i.e. the n-point correlation functions for n odd
are zero. The 
same property translates to the spherical harmonics space where the
expectation of even combinations of the coefficients $a_{lm}$ can be
expressed in terms of the power spectrum $C_l$ and the expectation 
of odd combinations, e.g. the bispectrum, is zero.
More generally, a very useful characteristic of a IGRF is that the
expectations of many 
statistical quantities can be 
calculated (semi)analytically. This is the case for the
morphological descriptors, number, shape and correlation of peaks
(maxima), 
scalars on the sphere, ...  

\begin{figure}
\centering
\includegraphics[height=7.1in, width=5.in]{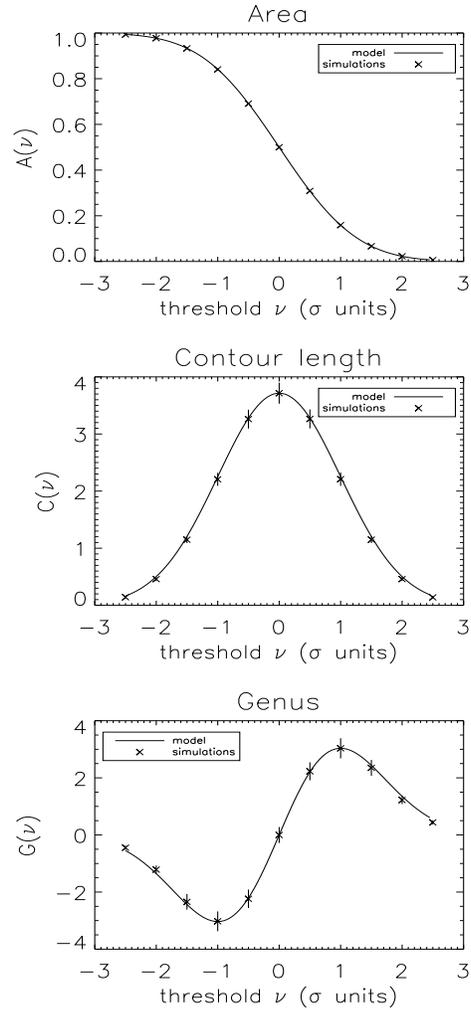}
\caption{Expected value of the Minkowski functionals for an
  IGRF obtained from eq.~\ref{eq:mf_c}, eq.~\ref{eq:mf_a},
  eq.~\ref{eq:mf_g}. Also 
  plotted are the average values (asterix) and error bars 
  of the Minkowski functionals obtained with 1000 Gaussian simulations. Taken
  from \cite{curto_07}.}
\label{fig:minkowski}       
\end{figure}

The Minkowski functionals are useful descriptors of the morphology of
point sets or smoothed fields in spaces with arbitrary dimension $d$ (see
e.g. \cite{adler_81}). As stated in \cite{schmalzing_gorski_98}, under a
few simple requirements any morphological descriptor can be written as
a linear combination of $d+1$ Minkowski functionals. For the sphere
they are therefore 3; namely, total length of
the contour ${\it C}(\nu)$ of the excursion set, total area ${\it
  A}(\nu)$ of the excursion set   
and the genus ${\it G}(\nu)$ above a threshold $\nu$. Their average
value per unit area for a IGRF can be simply given by     
\begin{equation}
<{\it C}(\nu)>=\frac{1}{8\theta_c} \exp \left(-\frac{\nu^2}{2}\right) \
\ ,\\
\label{eq:mf_c}
\end{equation} 
\begin{equation}
<{\it A}(\nu)>=\frac{1}{2}-\frac{2}{\sqrt{\pi}}\int^{\nu/\sqrt{2}}_0
\exp \left(-x^2\right) dx \ \ ,\\
\label{eq:mf_a}
\end{equation} 
\begin{equation}
<{\it G}(\nu)>\frac{1}{(2\pi)^{3/2}\theta_c^2}\nu
\exp\left(-\frac{\nu^2}{2}\right) \ \ ,  
\label{eq:mf_g}
\end{equation} 
where $\theta_c=\big(-\frac{C(0)}{C''(0)}\big)^{1/2}$ is the coherence
angle of the random field which is defined by the ratio of the two-point
correlation function $C(\theta)$ to its second derivative at zero lag
(see Fig.~\ref{fig:minkowski}).

Properties of peaks in a 2D IGRF have been studied by
\cite{longuet-higgins_57}, \cite{bond_efstathiou_87} and
\cite{barreiro_97}. The 
local description of a peak involves the second 
derivative of the field along the two principal directions. The
curvature radii are defined in the usual way from the second
derivative of the temperature anisotropies $\Delta$ at the position of the
maximum: 
$R_1=[-\Delta''_1(max)/2\sigma_0]^{-1/2}$
$R_2=[-\Delta''_2(max)/2\sigma_0]^{-1/2}$, where $\sigma_0$ is the 
anisotropy rms and $\Delta''_i$ is the second derivative along the
principal direction $i$. The two invariant
quantities, Gaussian curvature $\kappa$ and eccentricity $\epsilon$
can be constructed from them:
\begin{equation}
\kappa =\frac{1}{R_1 R_2} \ \ ,\ \ \ \epsilon
=\left[1-\left(\frac{R_2}{R_1}\right)^2\right]^{1/2} \ \ \ .
\end{equation}  
It is straightforward to obtain the number of peaks on the celestial
sphere 
$N(\kappa,\epsilon,\nu)\, d\kappa \, d\epsilon \, 
d\nu$ with Gaussian curvature, eccentricity and threshold between
$(\kappa ,\, \kappa + d\kappa )$, $(\epsilon ,\, \epsilon +
d\epsilon)$ and 
$(\nu ,\, \nu + d\nu)$, in terms of the two spectral parameters $\gamma$
and $\theta_*$ that characterize the cosmological model: 
\begin{eqnarray}
\gamma =\frac{\sigma_1^2}{\sigma_0\sigma_2} \ \ , \ \ \ \theta_* =
2^{1/2}\frac{\sigma_1}{\sigma_2} \ \ ,    \\
\sigma_0^2 = C(0) \ \ , \ \ \ \sigma_1^2 = -2C''(0) \ \ , \ \ \
\sigma_2^2 = \frac{8}{3}C^{(iv)}(0) \ \ .  
\end{eqnarray}
The number of peaks above a threshold $\nu$, $N(\nu)$, can be
calculated from the previous differential quantity by integrating over
the whole parameter space for $\kappa$ and $\epsilon$ and over the
interval $(\nu , \infty )$ for $\nu$:
\begin{eqnarray}
N(\nu) & = & N_T \left( \frac{6}{\pi} \right) ^{1/2} \exp(-\nu^2/2)
  \left\{ \gamma^2
  (\nu^2-1)\left[ 1-\frac{1}{2} {\rm erfc} (\gamma\nu s)\right] +
    \right. \nonumber   \\ 
& & + \nu\gamma (1-\gamma^2)\frac{s}{\pi^{1/2}}\exp(-\gamma^2\nu^2
  s^2) +  \\
& & \left. + t\left[1-\frac{1}{2} {\rm erfc} (\gamma\nu s
  t)\right]\exp(-\gamma^2\nu^2 t^2) \right\} \ , \nonumber
\end{eqnarray}
where
\begin{equation}
s=[2(1-\gamma^2)]^{-1/2} \ \ , \ \ \ t=(3-2\gamma^2)^{-1/2} \ \ ,
\end{equation}
and $N_T=(3^{1/2}\theta_*^2)^{-1}$ is the total number of peaks on
the whole celestial sphere.\\
Another interesting quantity is the distribution of the curvature of
the peaks. The pdf of peaks with inverse of the Gaussian curvature
$L\equiv 
\kappa^{-1}$ between $(L,\, L + dL)$ above the threshold $\nu$, $p(L$,
can be also obtained from $N(\kappa,\epsilon,\nu)\, d\kappa \, 
d\epsilon \, d\nu$:
\begin{eqnarray}
p(L) & = &
\left(\frac{6}{\pi}\right)^{1/2}(2\gamma\theta_c^2)^4tL^{-5}
\exp[(2\gamma\theta_c^2)^2L^{-2}] \nonumber \\ 
& & \int_\nu^\infty d\nu \exp(-3t^2\nu^2/2) {\rm
  erfc}\left[\frac{s}{t}(2\gamma\theta_c^2L^{-1}-\gamma\nu t^2\right]
\ . 
\end{eqnarray}
The distribution of excentricities can be calculated in a similar
manner. The pdf of peaks with eccentricity between $(\epsilon,\,
\epsilon + d\epsilon)$ above a threshold $\nu$, $p(\epsilon )$, can be
obtained after a straightforward calculation:
\begin{eqnarray}
p(\epsilon ) & = &
\frac{32\sqrt{6}}{\pi}\epsilon^3\frac{1-\epsilon^2}{(2-\epsilon^2)^5}
\nonumber \\
& & \int_\nu^\infty d\nu \exp(\nu^2/2) \left\{ (H\pi)^{1/2}
  \exp(-G)\left[ 1- {\rm erfc} (H^{1/2}\gamma\nu s)/2 \right]\times 
  \right. \nonumber \\
& &
\left[ 3H^2(1-\gamma^2)^2+6H^3\gamma^2(1-\gamma^2)\nu^2+(H\gamma\nu)^4\right]+
\nonumber \\
& & \left. \exp(-s^2\gamma^2\nu^2)s\left[ 5H^3\gamma
  (1-\gamma^2)^2\nu+H^4(\gamma\nu)^3(1-\gamma^2)\right] \right\} \ , 
\end{eqnarray}
where $H=(2-\epsilon^2)^2/[(3-2\gamma^2)\epsilon^4+4(1-\epsilon^2)]$ and
$G=H(\gamma\nu\epsilon^2)^2/(2-\epsilon^2)^2$.    

Scalar quantities can be constructed from the derivatives of the CMB
field on the sphere. A single scalar can be
constructed in terms of the ordinary derivative of the field
$\Delta_{,i}$. Only two independent scalars can be obtained from the
second covariant derivatives on the sphere, $\Delta_{;ij}$. Following
\cite{monteserin_05}, many scalar
quantities can be defined from the first and second 
derivatives of the field, associated to the Hessian matrix, the
distorsion, the gradient and the curvature. However, all except three
are correlated. For testing Gaussianity, it is convenient to use
normalized scalars for which the dependence of the scalars on 
the power spectrum has been removed. Here, as example, we focus on
three independent 
normalized scalars, namely the Laplacian, the fractional anisotropy and 
the square of the modulus of the gradient. The first two scalars have
been proved to be very 
efficient as detectors of non-Gaussianity \cite{monteserin_06}. The
third scalar, the square of the modulus of the 
gradient, is the only scalar from the list given in that paper which
is independent from all the others. It depends only on the first
derivatives of the field. The Laplacian and the fractional
anisotropy depend only on 
second derivatives and can be defined in terms of the eigenvalues
$\lambda_1$, $\lambda_2$, of the negative Hessian matrix ${\bf A}$ of the
field: ${\bf A}=(-T_{;ij})$. The eigenvalues, i.e. the negative
second derivatives along the two principal directions, can be written
as a function of the covariant second derivatives of the field
$\Delta({\bf n})$:
\begin{equation}
\lambda_1 =-\frac{1}{2}\left[
  \left(\Delta^{;i}_{~~i}\right)-\sqrt{\left(\Delta^{;i}_{~~i}\right)^2-2\left(
  \Delta^{;i}_{~~i}\Delta^{;j}_{~~j} -
  \Delta^{;j}_{~~i}\Delta^{;i}_{~~j} \right)} \right] \ ,
\end{equation}      
\begin{equation}
\lambda_2 =-\frac{1}{2}\left[
  \left(\Delta^{;i}_{~~i}\right)+\sqrt{\left(\Delta^{;i}_{~~i}\right)^2-2\left(
  \Delta^{;i}_{~~i}\Delta^{;j}_{~~j} -  
  \Delta^{;j}_{~~i}\Delta^{;i}_{~~j} \right)} \right] \ .
\end{equation}
We will assume $\lambda_1\ge\lambda_2$. Considering the values of
$\lambda_1$ and $\lambda_2$, three types of points can be
distinguished (see e.g. \cite{dore_03}: hill (both positive),
lake (both negative) and saddle (one positive and one negative)  \\      
The normalized Laplacian, or trace of the Hessian matrix,
$\bar{\lambda}_+$, 
is defined in terms of the eigenvalues as:
\begin{equation}
\bar{\lambda}_+= \frac{\lambda_1 + \lambda_2}{\sigma_2} \ , -\infty <
\bar{\lambda}_+ < \infty \ .
\end{equation} 
Since the Laplacian is given by linear transformations of the CMB
temperature fluctuation field $\Delta$, if the field is Gaussian then
its 1-point pdf is also Gaussian:
\begin{equation}
p(\bar{\lambda}_+)=\frac{1}{\sqrt{2\pi}}
\exp\left(-\frac{\bar{\lambda}_+^2}{2}\right) 
\label{eq:laplacian}
\end{equation} 
The fractional anisotropy \cite{basser_pierpaoli_96} has been used in
other fields like in the analysis of medical images. The normalized
quantity $\bar{f}_a$ is defined as: 
\begin{equation}
\bar{f}_a =
\frac{1}{\sqrt{2}}\frac{\bar\lambda_1^2-\bar\lambda_2^2}
     {\sqrt{\bar\lambda_1^2+\bar\lambda_2^2}}  
\end{equation}
where $\bar\lambda_1$ and $\bar\lambda_2$ are the normalized
eigenvalues given by:
\begin{equation}
\left(
\begin{array}{c}
\bar\lambda_1 \\
\bar\lambda_2
\end{array}
\right)
= 
\frac{1}{2}
\left(
\begin{array}{cc}
\frac{1}{\sigma_2}+\frac{1}{\sqrt{\sigma_2^2-2\sigma_1^2}} &
\frac{1}{\sigma_2}-\frac{1}{\sqrt{\sigma_2^2-2\sigma_1^2}} \\
\frac{1}{\sigma_2}-\frac{1}{\sqrt{\sigma_2^2-2\sigma_1^2}} &
\frac{1}{\sigma_2}+\frac{1}{\sqrt{\sigma_2^2-2\sigma_1^2}} 
\end{array}
\right)
\left(
\begin{array}{c}
\lambda_1 \\
\lambda_2
\end{array}
\right) \ .
\end{equation}
The pdf of the normalized fractional anisotropy is given by:
\begin{equation}
p(\bar{f}_a) = \frac{2\bar{f}_a}{(1-\bar{f}_a^2)^{1/2}
  (1+\bar{f}_a^2)^{3/2}} \ , 0<\bar{f}_a<1 \ \ .
\label{eq:fractional_anisotropy}
\end{equation}
The normalized square of the modulus of the gradient $\bar g$, which
depends only on the first derivatives of the field, is defined as:
\begin{equation}
\bar{g} = \frac{|\nabla \Delta |^2}{\sigma_1^2} =
\frac{\Delta^{,i}\Delta_{,i}}{\sigma_1^2} \ , 
\end{equation} 
where $\sigma_1^2$ is the dispersion of the unnormalized square of the
modulus of the gradient and
accounts for the normalization factor. 
In terms of the derivatives of the field with respect to the spherical
coordinates ($\theta,\phi$), $\bar g$ takes the form:
\begin{equation}
\bar{g} = \frac{1}{\sigma_1^2} \left[ \left(\frac{\partial
    \Delta}{\partial \theta}\right)^2 + 
\frac{1}{\sin^2\theta} \left(\frac{\partial \Delta}{\partial
  \phi}\right)^2 \right] \ . 
\end{equation}
Taking into account that the square of the gradient modulus is given
by the 
addition of two independent squared Gaussian variables, its pdf
follows a $\chi^2_2$ distribution. Since we consider the normalized
quantity 
$\bar g$, then its mean and dispersion are equal to one and its
distribution takes the simple form:
\begin{equation}
p(\bar{g}) = \exp(-\bar{g}) \ , \ \ \ 0 <\bar{D}_g < \infty \ .
\label{eq:gradient} 
\end{equation}

\begin{figure}
\centering
\includegraphics[height=7.in, width=3.in]{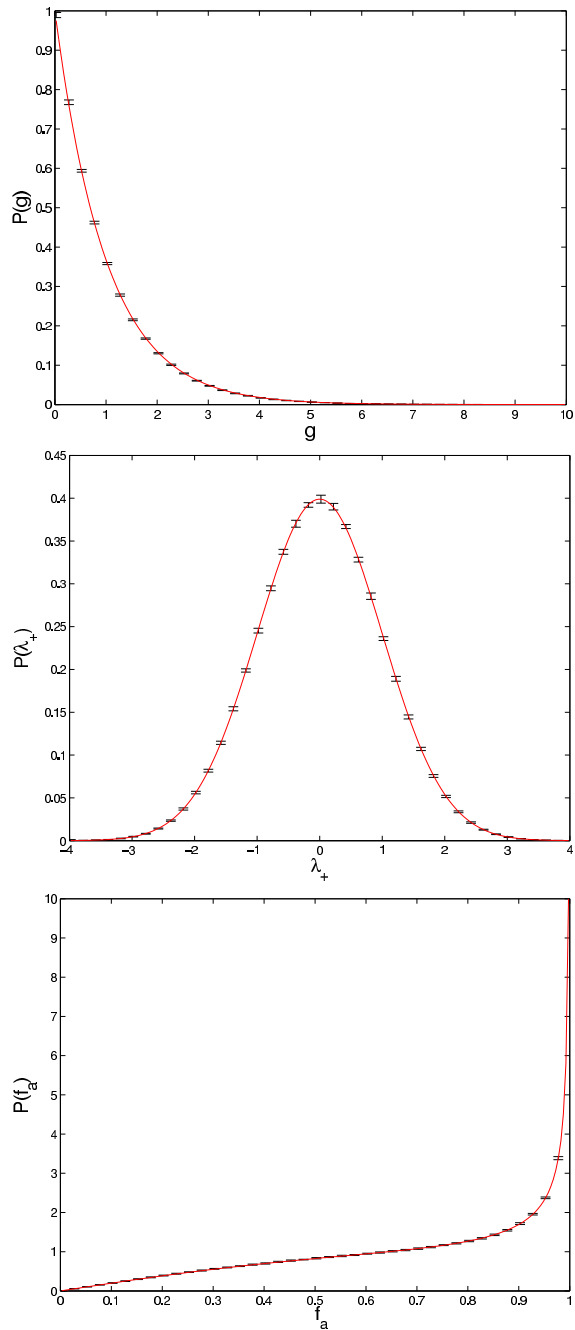}
\caption{Pdfs of the three normalized scalars described in the text:
  the Laplacian (eq.~\ref{eq:laplacian}), the fractional anisotropy
  (eq.~\ref{eq:fractional_anisotropy}) and  
the square of the modulus of the gradient (eq.~\ref{eq:gradient}). The
  average values and 
error bars of 1000 IGRF simulations are also represented. From
  \cite{monteserin_05}.} 
\label{fig:scalars_distributions}       
\end{figure}

\begin{sidewaysfigure}
\centering
\includegraphics{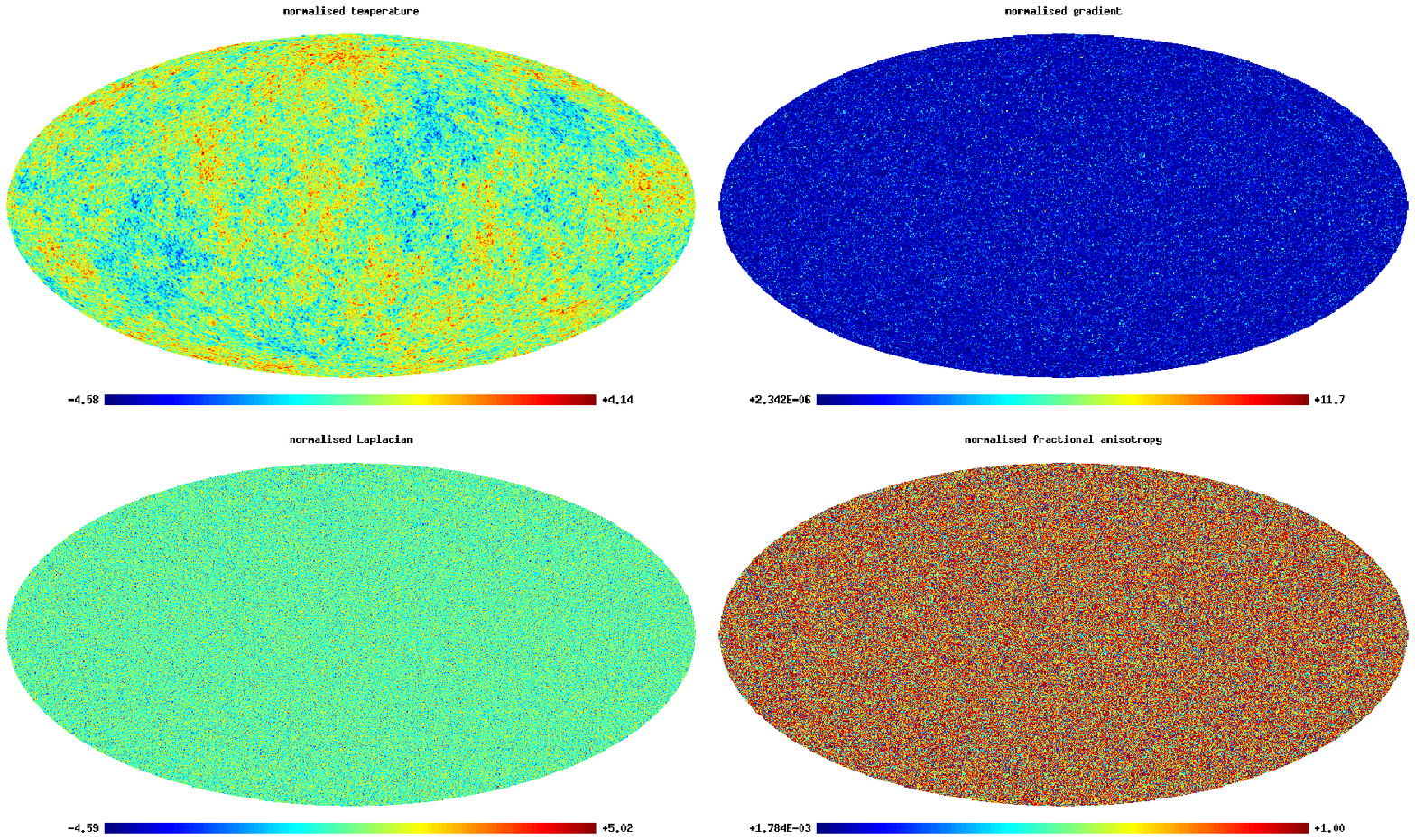}
\caption{Maps of the three normalized scalars described in the text
  and of the corresponding 
  normalized temperature for a realization of the IGRF. From
  \cite{monteserin_05}.} 
\label{fig:scalars_maps}       
\end{sidewaysfigure}

The pdfs of the three normalilzed scalars can be seen in
Fig.~\ref{fig:scalars_distributions}. Maps of the same scalars for a
random realization of a IGRF are shown in
Fig.~\ref{fig:scalars_maps}.

\section{Physical effects producing deviations from the standard
  IGRF \label{sec:physical_effects}} 

There are a number of physical effects which may produce deviations from
the standard IGRF. They are normally related to secondary
anisotropies produced by photon scattering, gravitational effects
generated by
the non-linear evolution of the matter density, variations from
standard inflation in the early universe, topological defects,
non-standard geometry and topology of the universe or primordial
magnetic fields. Below we sumarize some relevant aspects of the most  
studied effects.
 
\subsection{Secondary anisotropies \label{subsec:secondary_anisotropies}}

On the last scattering surface CMB anisotropies are generated via the
gravitational potential (Sachs-Wolfe effect, \cite{SW}) and the physics of the
baryon-photon plasma. These anisotropies are usually refered to as
{\sl primary anisotropies}. After matter-radiation decoupling when the
temperature drops below $\approx 3000K$, new
anisotropies are generated during the trip made by the CMB photons
to reach us. These {\sl secondary anisotropies} can be originated
via the gravitational redshift suffered by the photons when crossing
the gravitational potentials produced by the large-scale matter 
distribution, or by the scattering of the microwave photons with the
ionised matter after the reionization epoch ($z\lea 10$). The
gravitational potential can produce two types of effect: the gravitational
redshift suffered by the photons when they cross the varying potential
wells formed by the matter evolution (Rees-Sciama effect,
\cite{RS, martinez-gonzalez_sanz_90}), and the lensing effect produced
by the same  
gravitational potentials which bends their trayectory (see e.g. Bartelmann and
Schneider 2001 for a review). The CMB power spectrum produced by the
Rees-Sciama effect is subdominant on 
all angular scales \cite{sanz_96, hu_dodelson_02}, and the expected 3-point
correlations have an amplitude much below the cosmic variance
\cite{mollerach_95}. The lensing effect produces a smoothing of
the acoustic peaks in the CMB power spectrum \cite{seljak_96,
martinez-gonzalez_97}. It also induces high-order correlations in the CMB 
temperature and polarization fields \cite{hu_00}. 

The scattering of the CMB photons with the ionised medium produces a
randomisation of the directions of the rescattered photons,
implying a
supression of the primary CMB anisotropies whose resulting power
spectrum depends on the 
Thompson optical depth $\tau$ as $\exp(-2\tau)$. In addition, new anisotropies
are generated by the scattering with the 
free electrons moving along the line of sight which produces a Doppler
effect. This effect is strongly suppressed due to the
cancellation of velocities along the line of sight. However the
Doppler effect can survive cancellation if it is modulated by either
density or ionization fluctuations. Its amplitude is given by:
\begin{equation}
\frac {\Delta T}{T}=-\sigma_T\int dt e^{-\tau(\theta,t)}
n_e(\theta,t)v_r(\theta,t) \ \ .
\end{equation}     
In this formula $\sigma_T$ is the Thompson cross-section, $n_e$ the
electron density and $v_r$ the line of sight 
velocity of the electrons. $n_e$ can be further expressed as
$n_e(\theta,t)=\bar n_e(t) [1+\delta(\theta,t)+\delta_i(\theta,t)]$
where $\bar n_e(t)$ is 
the mean electron density, $\delta$ the density fluctuation and
$\delta_i$ the ionisation fraction. Two second order effects
generating secondary anisotropies appear in this formula. The first
one is the Doppler effect modulated by the density variation and
known as {\sl Ostriker and Vishniac effect} in the linear regime
\cite{OV, vishniac_87}. The second effect is the Doppler effect
modulated by the ionisation fraction and is usually referred to as
{\sl patchy reionization} (see e.g. \cite{aghanim_96}). Since they
are of second order, both effects produce non-Gaussian perturbations in
the CMB. However, their amplitude is much smaller than the thermal and
kinetic {\sl Sunyaev-Zeldovich effects (SZ)} \cite{SZ}
generated in the non-linear regime at lower redshifts and produced by
the scattering of hot, ionised gas associated to collapsed
structures. In addition to temperature anisotropies, due to the
primary quadrupole moment reionisation also produces polarisation. 

For more details about secondary anisotropies the reader
is refered to e.g. \cite{hu_dodelson_02} and \cite{aghanim_puget_06}.   

\subsection{Non-standard models of the early universe
  \label{subsec:non-standard_models}} 

\subsubsection{Non-standard inflationary models \label{subsubsec:inflation}}

In the standard, single field, slow roll inflationary model, the
dominant linear effects in the 
evolution of density fluctuations preserve the initial Gaussian
distribution produced by the quantum fluctuations of the inflaton. 
On the contrary, second-order effects perturb the original Gaussianity 
of the fluctuations, although resulting in a negligible effect in the
standard model.   
In non-standard inflationary models, primordial non-Gaussianity should
be added to the second-order effects associated to the evolution of 
density fluctuations after inflation finishes. In any case, it is
common to 
characterize phenomenologically the deviations from Gaussianity by 
introducing the $f_{NL}$ parameter in the gravitational potential
\cite{komatsu_spergel_01}: 
\begin{equation}
\phi = \phi_L + f_{NL} \phi_L^2\ \ ,
\label{eq:f_NL}
\end{equation}
where $\phi_L$ is the gravitational potential at the linear order and it is 
distributed following a Gaussian random field. In general the
{\sl non-linear couplig parameter} $f_{NL}$ is a function of the distance
vectors and the product  
is really a convolution. However, an effective $f_{NL}$ which accounts
for those complexities can still be used. A detailed 
study of the values of $f_{NL}$ for several non-standard inflationary
models, including multi-field inflation, the curvaton scenario,
inhomogeneous 
reheating and the Dirac-Born-Infeld (DBI) inflation inspired by string
theory, can be found in \cite{bartolo_04} (see 
also \cite{matarrese_08} in this volume). Here it is worth 
noticing that sometimes the specific
$f_{NL}$ parameter given by eq.~\ref{eq:f_NL} is called the {\sl local}
non-linear coupling parameter
$f^{local}_{NL}$, refering to the fact that $\phi$ is obtained from
$\phi_L$ at the same position in space. In addition to
$f^{local}_{NL}$ which contains mainly information of the squeezed
configurations (those with two large and similar wave vectors and the
other small),
the {\sl equilateral} non-linear coupling parameter 
$f^{equil}_{NL}$ is used to caracterize equilateral configurations of
the bispectrum for which the lengths of the three wave vectors are
equal in Fourier space. The two non-linear coupling parameters suppose
a fair representation of a large class of models. For instance,
$f^{local}_{NL}$ can be generated in curvaton and reheating scenarios
whereas $f^{equil}_{NL}$ can be produced in DBI inflation within the
context of String theory. 

In the standard single-field slow-roll inflation the effective
$f_{NL}$ is dominated by second-order gravitational corrections
leading to values of order unity. These low values require very 
sensitive measurements to be detected and are not even within reach of
the Planck mission. Non-standard inflationary models generally predict
larger values, some of them already constrained by WMAP. A positive
detection of $f_{NL}\gsim 10$ would rule out the standard inflationary
models.   

For an ideal experiment with white noise, no
foreground residuals and no Galactic and point source mask, it has
been shown that the
optimal estimator for $f_{NL}$ is based on a bispectrum test
constructed from a cubic combination of appropriately filtered
temperature and 
polarization maps \cite{komatsu_05, creminelli_06, yadav_07a}. An
extension to this estimator to deal with data under 
realistic experimental conditions has been made by
\cite{yadav_07b}. Several observational constraints on $f_{NL}$ have
been  
derived from different experiments. The first ones were obtained with
COBE-DMR \cite{komatsu_02, cayon_03}. Further constraints
have been derived with other experiments like VSA \cite{smith_04},
BOOMERANG \cite{troia_07} and Archeops \cite{curto_07,
  curto_08}. However, the best limits have been derived with the 
various releases of the WMAP data \cite{komatsu_03,
spergel_07, komatsu_08} representing an improvement of at least an order 
of magnitud over previous ones (see below).           

\subsubsection{Topological defects \label{subsubsec:topological_defects}}

In standard theories of particle physics the fundamental forces of
nature unify progressively when the energy scale exceeds certain
thresholds. These unified theories imply that the universe went
through several phase transitions during the early stages of its
evolution. At energies above $10^2$GeV, the electromagnetism and the weak
nuclear force are merged into the electroweak force. At 
higher energies of the order of $10^{15}$GeV it is believed that the
electroweak force unifies with the strong nuclear force, a process
usually called Grand Unified Theory (GUT). At even higher energies it is
speculated that it is possible to merge gravity with the other three
interactions. 

Unified theories are based on symmetry. When the symmetry is broken
spontaneously via the Higgs mechanism, topological defects generically
appear \cite{kibble_76} (for a more pedagogical discussion about
topological defects see \cite{vilenkin_shellard_00}). Depending on the 
dimensionality of the symmetry which is broken, a type of topological 
defect is formed. When a discrete symmetry is broken domain walls
form, in a two-dimensional symmetry breaking cosmic strings appear,
monopoles form when the symmetry breaking is three-dimensional and
textures appear when it is in four or more dimensions. In the 1980s
topological defects were considered as an alternative scenario to
inflationary quantum fluctuations for the process of structure formation.
CMB observations showed that the former scenario could only play a
subdominant role as the source of cosmic structure. Moreover, due to
the catastrophic effects that the presence of domain walls and
monopoles would have on our universe only the existence of cosmic
strings and textures is usually considered.

A lot of effort has been made to study the cosmological consequences
of cosmic strings. It is generally believed that the best strategy to
detect 
them is by searching for their imprint on the CMB maps. The
non-Gaussian character of the density field of 
strings produce line discontinuities in the CMB anisotropies 
at arcmin angular scales as a consequence of the metric deformation
around the strings (the {Kaiser-Sttebins effect}
\cite{kaiser_stebbins_84}). However, at 
larger angular scales a Gaussian distribution emerge from the central
limit theorem. 

Cosmic textures were first studied in detail by \cite{turok_89}. They are
much more diffuse than the other defects which are localized at a
point (monopole), on a line (cosmic string) or a surface (domain
wall). Contrary to the others, they are unstable and consist of
twisted configurations of fields which collapse and unwind. Each
texture creates a time-varying gravitational potential which produces
red- or blue-shift to the CMB photons passing through such a region.
Thus, textures generate hot and cold spots on the CMB anisotropy
maps whose amplitude is set by the symmetry breaking energy scale. The
shape of the spots is approximately spherically symmetric and
an approximated analytical formula has been derived by
\cite{turok_spergel_90}. Recently, a very cold spot detected in the WMAP 
temperature map has been found to be consistent with the effect
produced by a texture \cite{cruz_07}. If confirmed, this result
will have outstanding consequences in our understanding of the
universe (see Sec.~\ref{subsubsec:texture} for more details on this finding). 

\subsection{Non-standard geometry and topology
  \label{subsec:geometry_topology}} 

The geometry of the observable universe is believed to be well
approximated on large 
scales by a Friedmann-Lemaitre-Robertson-Walker (FLRW) metric characterized
by a homogeneous and isotropic space-time. Support for the homogeneity
of the universe is provided by the largest surveys of the galaxy
distribution (e.g. SDSS DR6 \cite{adelman-mccarthy_08}, 2dFGRS
\cite{colless_01}) and by the smallness of the CMB
temperature fluctuations. The cosmological principle, stating that the
universe is homogeneous and isotropic on large scales, follows from
those observations and the assumption that we are not located at a
special place in the universe (the Copernican principle).

However, recent observations of the CMB as measured by the
WMAP satellite might question the cosmological principle. They
are the large scale power asymmetry found between the two hemispheres of a
reference frame close to that of the ecliptic one \cite{eriksen_04a,
hansen_04a}, the planarity and alignment of the 
low multipoles \cite{oliveira-costa_04, land_magueijo_05a}, the
non-Gaussian cold spot present at   
scales of $\approx 10^\circ$ \cite{vielva_04, cruz_05} and
the alignment of local structures also at similar scales
\cite{wiaux_06, vielva_07a} (see Sec.~\ref{sec:WMAP_anomalies} for more 
details). These results may imply the existence of priviledged
directions in the CMB map and thus motivates the study of alternatives
to the standard FLRW metric. 

An interesting class of alternative models is that for which the metric is
homogeneous but anisotropic. They are known as Bianchi models and are
classified acording to their space-time properties. Their predictions
for the CMB anisotropy were studied in \cite{barrow_85}. Since  
the signatures left by those models appear on large angular scales,
they have been already constrained by the COBE-DMR experiment data
\cite{martinez-gonzalez_sanz_95, bunn_96, kogut_97}. One 
particularly interesting case is Bianchi VII$_h$ which experiences
anisotropic expansion and global rotation. These two properties
produce a characteristic pattern in the CMB in the form of a spiral
pattern and spots. This model has been recently used to account for
some of the large scale WMAP anomalies mentioned above \cite{jaffe_05,
  jaffe_06a, jaffe_06b, bridges_07a} (see Sec.~\ref{subsubsec:bianchi} for more
details). An important problem with 
this model comparison approach is that the CMB anisotropies are not
computed in a self-consistent way but are assumed to be the sum of two
independent components: an isotropic one produced by the
energy-density fluctuations which is assumed to be the same than for
the FLRW model, and an anisotropic component which is the determinisitc effect
produced by the anisotropic model.

A different source of anisotropic features in the CMB sky is the
global topology of the universe. The local character of the General
Theory of Relativity does not theoretically constrain it. If the
topology of the universe is non-trivial (i.e. multiconnected, meaning
that there is not a unique way to connect two points by geodesics)
then CMB photons originated from the same location on the
last scattering surface can be observed in different directions. This
effect manifests itself in the CMB sky as anisotropic patterns,
correlated (matched) circles \cite{cornish_04} or, more
generically, as deviations from a IGRF (for a discussion on different
topologies and tests developed to detect them see the reviews by
\cite{lachieze-rey_luminet_95} and \cite{levin_02}). An additional
consequence of a multiconnected universe is the lack of fluctuations
in the CMB above the wavelength corresponding to the size of the
universe. This property has led to several authors to suggest that the
low quadrupole and the alignment of the low multipoles measured by
WMAP might be an evidence of a non-trivial 
topology \cite{luminet_03, cresswell_06}. Constraints on
the topology of the universe started with the COBE-DMR data
\cite{oliveira-costa_96, roukema_00, rocha_04} and  
followed with the WMAP data \cite{inoue_sugiyama_03, kunz_06, kunz_08,
  phillips_kogut_06, niarchou_jaffe_07}. All 
those analyses concluded that the WMAP data do not show any evidence of
multiconnected universes.      
  
\subsection{Primordial magnetic fields}

In recent years there has been an increasing interest in studying the
consequences that the possible existence of primordial magnetic fields
might have 
on the CMB. Magnetic fields of order of a few $\mu$G have been measured in a
wide range of astrophysical structures, from individual galaxies
\cite{han_02} to galaxy clusters \cite{kronberg_04}. It is also widely
beleived that they are present in 
superclusters. Although the origin of those 
magnetic fields is still unclear and their existence does not
necessarily imply a primordial origin, studying the interplay between
magnetic fields and CMB 
is justified by the important consequences that it may have on the
CMB temperature and polarization anisotropy, 
and also on the distorsion of the blackbody spectrum (for reviews on
this topic the reader is referred to \cite{giovannini_04, durrer_07}).
In particular, the presence of magnetic fields introduce
non-Gaussianity in the CMB anisotropy 
since its amplitude depends on the square of the magnetic field intensity.

Two very different cases can be considered for the primordial magnetic
field: a uniform field and a
stochastic one. The former breaks the spatial isotropy of the
background geometry by introducing shear through an anisotropic stress. This
leads to the well known homogeneous and anisotropic Bianchi models
\cite{barrow_85}. Since those models also imply an anisotropic CMB
field, a uniform magnetic field generates phase
correlations between different $a_{lm}$. The uniform magnetic field
has been constrained with the CMB quadrupole and also with the phase
correlations, implying comparable constraints on the magnetic field
intensity of a few nG \cite{barrow_97, durrer_98, chen_04}.

The stochastic magnetic field is a more realistic scenario which can
be generated during inflation. In this case the isotropy of the
background geometry is preserved. Temperature and polarization anisotropies are
generated through the vector and tensor modes associated to the
magnetic field energy-momentum tensor. Allowed 
amplitudes of the magnetic field intensity of about several nG can 
produce a potentially observable B-mode polarization for nearly scale
invariant spectra \cite{lewis_04}. This signal could be distinguished
from the one generated by the inflationary gravitational wave background by its
non-Gaussian character. In addition Faraday rotation induces B-mode
polarization from the ordinary E-mode with the characteristic $\nu^{-2}$
dependence. 

Therefore, the presence of a primordial magnetic field can leave
unambiguous inprints in the CMB anisotropy that would allow its
identification with the sensitive data expected from the coming experiments.
      
\section{Methods to test Gaussianity \label{sec:methods}}

Testing the Gaussianity of CMB data is not an easy task. In principle
it consists in just proving the properties of the IGRF that we
discussed in Sec.~\ref{subsec:IGRF_properties}: isotropy and
multinormality. The CMB data 
represent a single realization of the underlying random field which
for the standard model is nearly Gaussian and isotropic. In practice
the analysis is complicated by the characteristics of the experiment
which need to 
be known very precisely: calibration uncertainties, instrumental noise
(white and 1/f) which is 
normally anisotropic in pixel space depending on the scanning
strategy, beam response (usually close to Gaussian), data processing,
... And by foreground contamination which demands certain
previous cleaning
operations in the data, requiring masking certain areas where the
foregrounds are very intense and leaving some amount of residuals 
in the rest of the surveyed area. The result of cleaning depends on 
our a priori knowledge of the physical properties of the foregrounds
and the component separation method used for their removal. These
ingredients must be considered in the analysis by performing
simulations accounting for them.

Unless one is interested in the compatibility of the data with a
specific alternative model for which and optimal method may be found
(as specific non-standard inflation, geometry or topology),
there are infinite ways in which a random field can deviate
from the IGRF one. Methods to test Gaussianity can be classified
by the type of property that they try to prove. Typical examples are 
cumulants and n-point correlation functions in real space (the former
should vanish and the latter either vanish for the odd order or can be 
expressed in terms of 2-point correlation functions for the even
order), moments in spherical harmonics space 
(bispectrum, trispectrum, ...), or moments in other spaces to which the
data is transformed by linear operations which preserve 
Gaussianity: filters, wavelets, signal-to-noise eigenvectors, ...
Other approaches may test different properties of the CMB random
field, like the morphology of the data using the
Minkowski Functionals or the geometry of the data using scalar
quantities constructed by the first and second covariant derivatives
as, for example, the local curvature (see
Sec.~\ref{subsec:IGRF_properties}). These and other methods are described in 
more detail in \cite{martinez-gonzalez_08} and \cite{barreiro_07} and
references therein. As 
example of the typical statistical procedure followed in the analysis,
below we focus on a few statistical methods which have been
often used in the literature based on the Minkowski Functionals,
bispectrum or wavelets.

As we have described in Sec.~\ref{subsec:IGRF_properties}, for the
excursion set above a 
given threshold $\nu$, there are three Minkowski
Functionals on the sphere: total contour lenght ${\it C}(\nu)$, total
area ${\it A}(\nu)$ and the 
genus ${\it G}(\nu)$ (see \cite{schmalzing_gorski_98}). In the case of
an IGRF their 
expected values follow simple analytical expresions as a function of
$\nu$ (eqs.~{\ref{eq:mf_c},\ref{eq:mf_a},\ref{eq:mf_g}}). However,
simulations are needed to account for the 
experimental characteristics, basically noise, beam response and
mask. As it can be shown with simulations, the 1-pdf for any Minkowski
Functional at 
each $\nu$ follows a nice bell-shape distribution, implying the natural
choice of a generalized $\chi^2$ as the appropriate statistical test
to be used in this case to combine all the information. More specifically,
considering $n_{th}$ different thresholds we can define a $3n_{th}$
vector $\bf v$, 
\begin{equation}
{\bf v} = \left({\it A}(\nu_1),..,{\it A}(\nu_{n_{th}}),{\it
  C}(\nu_1),..,{\it C}(\nu_{n_{th}}),{\it G}(\nu_1),..,{\it
  G}(\nu_{n_{th}})\right) \ \ .
\end{equation}
The generalized $\chi^2$ statistic to test the Gaussianity of a data
map can be then constructed as
\begin{equation}
\chi^2 = \sum^{3n_{th}}_{i,j=1} \left(v_i-<v_i>\right) C_{ij}
\left(v_j-<v_j>\right)
\label{eq:chi2} 
\end{equation}
where $<>$ is the expected value for the Gaussian case and $C_{ij}$ is
the covariance matrix, $C_{ij}=<v_iv_j>-<v_i><v_j>$, both of them usually
constructed with simulations. For testing the compatibility of the
data with a non-Gaussian model (e.g. a non-standard inflationary model
characterized by the $f_{NL}$ parameter) we simply have to use the
corresponding expected value and covariance matrix for the Minkowski
Functionals at different thresholds in eq.~\ref{eq:chi2}. If the
deviations from 
Gaussianity are small (as in the case of the $f_{NL}$ models with
$f_{NL}\lea 1000$) the covariance
matrix can be well aproximated by the one of the Gaussian case.      
More information is added to the analysis by considering $n_{res}$
different resolutions of a given data map. Including this extra
information simply increases the vectors and covariance
matrix present in the $\chi^2$ expression to a dimension
$3n_{th}n_{res}$. Examples of applications of this method to different
data sets can be found in \cite{komatsu_03} for the 1-year WMAP
data, \cite{troia_07} for the BOOMERanG 3-year data and
\cite{curto_08} for the Archeops data. Recently, perturvative formulae of the
Minkowski Functionals as a function of $f_{NL}$  
have been derived for the $f_{NL}$ models \cite{hikage_06}. The results of
applying them to the 3-year WMAP data show constraints on $f_{NL}$ very
similar to the ones obtained with the optimal bispectrum \cite{hikage_08}.  

Generically, non-standard models of inflation produce small deviations
of Gaussianity which are more prominent in the 3-point correlation
function or equivalently, its harmonic transform the bispectrum $B^{m_1
  m_2 m_3}_{l_1 l_2 
  l_3} \equiv <a_{l_1 m_1}a_{l_2 m_2}a_{l_3 m_3}>$. The trispectrum,
the harmonic transform of the 4-point correlation function, 
can also play an important role in discriminating inflation models
since some models do not produce any bispectra but produce significant
trispectra, or produce similar amplitudes of the bispectra but very
different trispectra (e.g. DBI inflation, \cite{huang_shiu_06}, or
New Ekpyrotic Cosmology, \cite{buchbinder_07}). Here we
briefly describe the bispectrum, for more details on it and the
trispectrum see \cite{komatsu_01}. The average
bispectrum, $B_{l_1 l_2 l_3}$, is the 
rotationally invariant third order moment of spherical harmonic
coefficients and is given by the following expresion (see \cite{hu_01} and
\cite{bartolo_04} for more details on this and higher order
moments):
\begin{equation}
B_{l_1 l_2 l_3}=\sum_{m_1 m_2 m_3} 
\left(
\begin{array}{ccc}
l_1 & l_2 & l_3 \\
m_1 & m_2 & m_3
\end{array}
\right)
<a_{l_1 m_1}a_{l_2 m_2}a_{l_3 m_3}> \ \ ,
\label{eq:averaged_bispectrum}
\end{equation}
where $(...)$ is the Wigner-3j symbol. The bispectrum must satisfy the
selection rules. Rotational invariance of the 3-point correlation
function implies that the bispectrum can be written as
\begin{equation}
B^{m_1 m_2 m_3}_{l_1 l_2 l_3}= G^{m_1 m_2 m_3}_{l_1 l_2 l_3}  b_{l_1
  l_2 l_3} \ \ ,
\label{eq:reduced_bispectrum}
\end{equation}
where $G^{m_1 m_2 m_3}_{l_1 l_2 l_3}$ is the Gaunt factor and $b_{l_1
  l_2 l_3}$ 
is a real symmetric function of $l_1$, $l_2$ and $l_3$ called the
{\sl reduced bispectrum}. By substituting eq.~\ref{eq:reduced_bispectrum} in
eq.~\ref{eq:averaged_bispectrum} it is straight forward to obtain the following
relation between the averaged bispectrum and the reduced bispectrum
\begin{equation}
B_{l_1 l_2 l_3}= \sqrt{\frac{(2l_1+1)(2l_2+1)(2l_3+1)}{4\pi}} 
\left(
\begin{array}{ccc}
l_1 & l_2 & l_3 \\
0 & 0 & 0
\end{array}
\right)
b_{l_1 l_2 l_3} \ \ .
\end{equation} 
Therefore under rotational invariance the reduced bispectrum contains
all physical information of the bispectrum. In particular $b_{l_1 l_2
  l_3}$ is fully identified by the inflationary models characterized
by the non-linear coupling parameter $f_{NL}$. Thus these models can be
optimally tested by using the averaged bispectrum. The computation of
the full averaged bispectrum scales as $N^{5/2}_{pix}$, where $N_{pix}$ is the
number of pixels, and is already not feasible for the WMAP
data. \cite{komatsu_05} solved this problem by constructing a
cubic statistic, the KSW estimator, that combines the triangle
configurations of the 
bispectrum optimally for determining $f^{local}_{NL}$, with its
computation scaling as $N^{3/2}_{pix}$. The extension of this estimator
to $f^{equil}_{NL}$, and also including polarization, has been made in
\cite{creminelli_06} and \cite{yadav_07a}. The strongest
constraints on the non-linear coupling parameter up to date have been
obtained by applying the KSW estimator to the WMAP data (see next section).

Wavelets are compensated filters which allow one to extract
information which is localized in both real and harmonics
space. In particular wavelets may provide information on the position
and scale of different features in astrophysical images. They
can be more sensitive than classical methods (for a review on wavelets
see \cite{jones_08} and for applications to the CMB see
\cite{vielva_07b}). One example which has been used many times in  
cosmological applications is the Mexican hat wavelet defined as the
Laplacian of a Gaussian function. For CMB analyses the extension of an
Euclidean wavelet to the sphere can be made with an inverse stereographic
projection \cite{antoine_vandergheynst_98}. By this procedure the
spherical Mexican hat wavelet can be constructed preserving dilation and
compensation \cite{martinez-gonzalez_02}, and was first applied to
CMB analyses by \cite{cayon_01} using COBE-DMR data. Also it can be made
directional, i.e. sensitive not only to the scale but also to the
orientation of a feature, by simply considering different widths along
the two axes of the original 2-dimensional Gaussian.\\
A very useful property to study directional properties of structures in
an image is the so called steerability. In general, a directional or
non-axisymmetric filter is said to be {\sl steerable} 
if any rotation about itself can be expressed as a finite linear
combination of non-rotated basis filters. This concept has been recently
extended to the sphere by \cite{wiaux_05}. An important consequence of
the steerability property is that it makes the wavelet analysis very
efficient computationally (see Sec.~\ref{subsec:local_structures}). An
interesting spherical steerable 
wavelet is the second Gaussian derivative one which may be rotated in
terms of three basis wavelets: the second derivative in direction x,
the second derivative in direction y, and the cross-derivative. This
wavelet provides information on the three local morphological measures
of orientation, signed intensity (amplitude at the orientation which
maximizes the absolute value of the coefficient) and elongation. It
has been recently applied to the CMB analysis to test global isotropy
and Gaussianity \cite{wiaux_06, wiaux_08, vielva_07a}
(see Sec.~\ref{subsec:local_structures}).      

\section{Constraints from observations \label{sec:observations}}

Previously to any Gaussianity study, one major problem in the analysis
of the observational data is the 
separation of the different Galactic and extragalactic components from
the CMB itself. A key property to distinguish them is the frequency
dependence of the specific emission in the microwave range. Thus, the
intensity of the 
Galactic synchrotron and free-free emissions decreases with frequency
as a power-law with approximated spectral indexes $3$ and $2$,
respectively. On the contrary, the intensity of the Galactic thermal
dust increases with frequency following approximately a grey-body
spectrum with emissivity $\propto \nu^2$ and temperature $T_D
\approx 10-20$K. Extragalactic sources emitting in the microwave band
are typically radio galaxies and IR galaxies, with their emission
dominating at frequencies $\lea 100$GHz for the former
(synchrotron-like) and at higher ones (dust-like) for the latter. An
additional extragalactic emission comes from galaxy clusters through
the SZ effect, as was discussed in
Sec.~\ref{subsec:secondary_anisotropies}. The 
spatial distribution of the different foregrounds is also very
different from the CMB one, differing very much from that of an
IGRF. Thus, the Galactic foreground emissions 
dominate at large angular scales with the power spectrum of
fluctuations decaying approximately as $C_\ell \propto \ell^{-3}$. On
the contrary the extragalactic foregrounds dominate at the smallest
angular scales and appear as point sources for typical CMB
experiments. (For more details on the properties of foregrounds see
contributions by \cite{davies_08} and \cite{partridge_08} to this
volume). All these properties have to be exploited in order to 
best disentangle the foreground emissions from the CMB one. Indeed,
the numerous component separation methods already developed take
advantage of our knowledge of those foreground properties to obtain a
clean CMB map (see \cite{delabrouille_cardoso_08} and \cite{barreiro_08} for
further reading on this topic).

The Gaussianity of the CMB signal has been studied with data measured
by many different experiments. The first systematic analysis was
carried out with the all-sky COBE-DMR satellite \cite{smoot_92,
bennett_94}. Other analyses were
based on data covering a fraction of the sky and obtained from
experiments onboard stratospheric balloons, like QMAP, MAXIMA,
BOOMERanG or Archeops, and from ground-based ones like Saskatoon,
QMASK or VSA. The result of most of those Gaussianity studies was a
systematic compatibility with the standard IGRF. Deviations were
also claimed in a few cases which were later proved to be due to
either systematics or an incomplete analysis. Upper limits on the
$f^{local}_{NL}$ parameter were derived of approximately a few thousands
\cite{komatsu_02, cayon_03}. For a more detailed 
discussion on those analyses see \cite{martinez-gonzalez_08}. A significant
improvement has been recently achieved with measurements by BOOMERanG
and Archeops, lowering the upper limit to $f_{NL}\lea 1000$ at the
$2\sigma$ c.l. \cite{troia_07, curto_08}. It is worth
mentioning a deviation from the IGRF found in one of the VSA fields,
the Corona Borealis supercluster \cite{genova-santos_05,
rubino-martin_06}. The deviation consists in a strong and
resolved negative spot, of $\approx -250\mu$K and angular size of
$\approx 20$ arcmin, which is not associated with any of the clusters
of that supercluster. A SZ effect produced by a
diffuse, extended warm/hot gas distribution has been suggested as a
possible explanation \cite{genova-santos_08}. This hypothesis,
if confirmed, would be of relevance for providing the location of the
missing baryons in the local universe.  

The precision with which the Gaussianity of the CMB can be tested has
been strongly improved with the quality data measured with the WMAP
satellite \cite{bennett_03a, hinshaw_07, hinshaw_08}. The WMAP
team has provided ``cleaned'' CMB maps at the Q, V and W frequency bands
for the 5 year data collected by the experiment, and masks covering a
region around the Galactic plane and a catalogue of radio sources
where the foreground emission cannot be removed at the required
sensitivity. By a noise-weighting combination of the cleaned maps at
the cosmological frequencies and applying a conservative mask the WMAP
team has 
performed a Gaussianity study of the data based on the Minkowski
Functionals and the optimal bispectrum. For both quantities the 1, 3
and 5-year WMAP 
data releases have been found to be compatible with the IGRF
\cite{komatsu_03, spergel_07, komatsu_08}. Stringent limits 
have been also derived on the local and equilateral $f_{NL}$ parameter
using an optimal bispectrum-like quantity: $-9<f^{local}_{NL}<111$ and
$-151<f^{equil}_{NL}<253$ at the $2\sigma$ c.l.
 
\section{WMAP anomalies \label{sec:WMAP_anomalies}}

The WMAP team found the WMAP data consistent with the IGRF using the
Minkowski functionals and the bispectrum. Subsequently, many works have
tested the Gaussianity of the WMAP data in many different ways. Some
of them have found agreement with Gaussianity whereas others found
significant deviations of the IGRF. Examples of the former are
analyses based on
3-point correlation analysis \cite{gaztanaga_wagg_03,
chen_szapudi_05}, integrated bispectrum \cite{cabella_06},    
real space statistics \cite{bartosz_08} or
isotropy analyses based on the Bipolar power spectrum \cite{hajian_05,
  hajian_souradeep_06}. Examples of the latter are 
analyses based on phase correlations \cite{chiang_03}, the genus
\cite{park_04}, isotropic wavelets \cite{vielva_04, mukherjee_wang_04,
  cruz_05}, 
1-pdf \cite{monteserin_07}, isotropy  
analyses based on local n-point correlations \cite{eriksen_04a}, local
curvature \cite{hansen_04b}, multipole vectors \cite{copi_04, copi_07}
or directional wavelets \cite{mcewen_05, wiaux_06, vielva_07a,
  wiaux_08}. However, some of the  
analyses have been performed on the whole sky internal linear
combination (ILC) map \cite{bennett_03b} or similar maps
(e.g. \cite{TOH}) which is well 
known to suffer from Galactic contamination and, as stated by the WMAP
team itself, should not be used for cosmological analyses. From here
on, we will concentrate on the most relevant works based on WMAP maps
where a certain region around the Galactic plane has been masked as
well as several hundred of extragalactic sources.
The deviations or ``anomalies'' reported 
have been detected using other statistical quantities different from the
ones originally used by the WMAP team. They are the North-South asymmetry
\cite{eriksen_04a, hansen_04a}, alignment of the low multipoles
\cite{oliveira-costa_04, land_magueijo_05a}, the cold spot
\cite{vielva_04, cruz_05,
cruz_07}, non-Gaussian features detected with directional
wavelets \cite{mcewen_05, mcewen_08}, 
alignment of CMB structures \cite{wiaux_06, vielva_07a}, low variance
\cite{monteserin_07}.
Below we describe these anomalies and discuss several relevant
aspects like significance, origin, etc.
  
\subsection{North-south asymmetry \label{subsec:asymmetry}}

\begin{sidewaysfigure}
\centering
\includegraphics[angle=-90, width=1\textwidth]{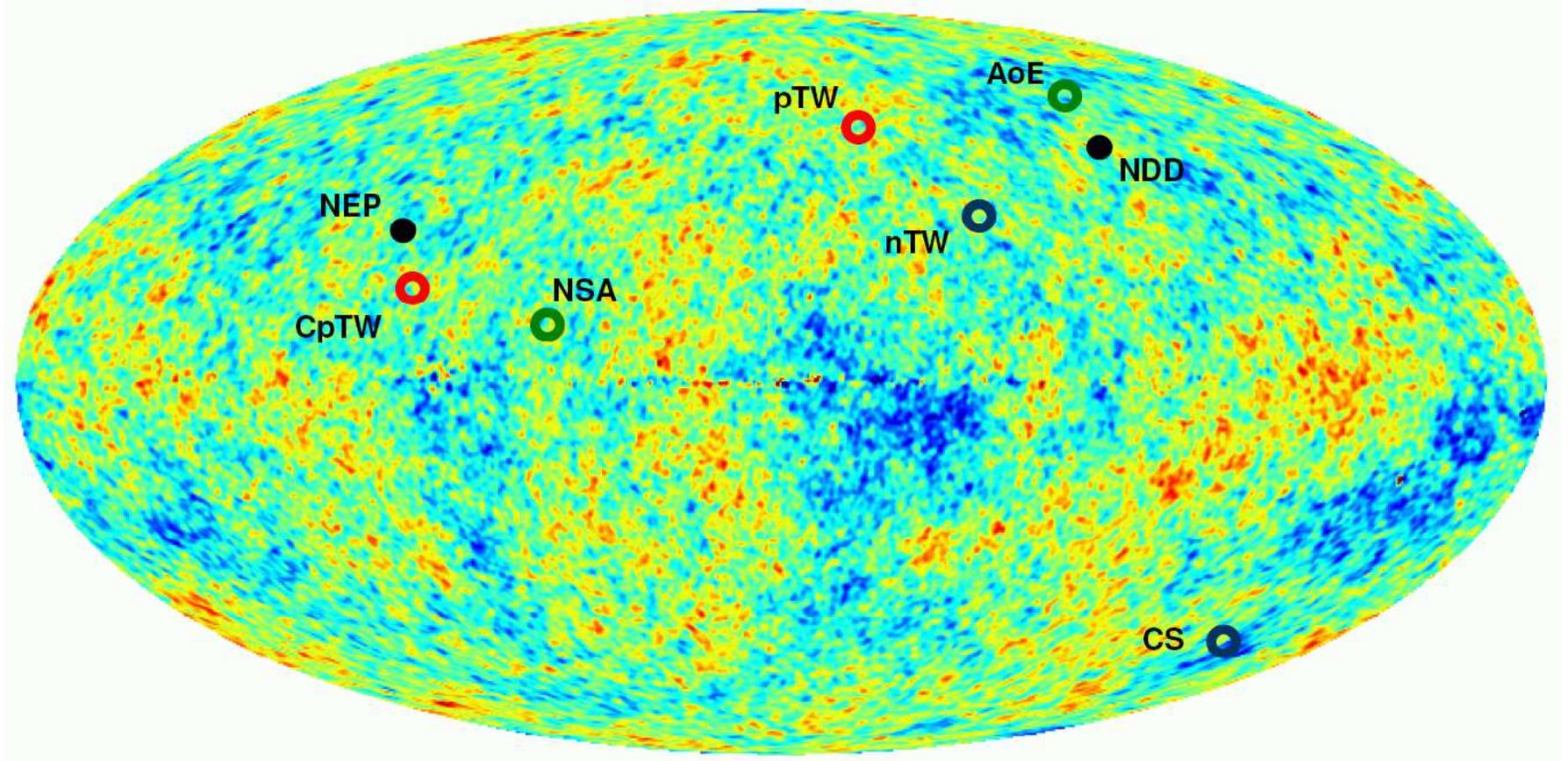}
\caption{Directions in the microwave sky derived from the anomalies
  found in WMAP data: northern direction of the North-South asymmetry
  (NSA), axis of 
  evil (AoE), the cold spot (CS), cluster of positive total weights
  (CpTW), perpendicular axis to the positive total weights plane
  (pTW), perpendicular axis to the negative total weights plane
  (nTW). For reference the North ecliptic pole (NEP) and the northern
  direction of the CMB dipole (NDD) are also shown.}
\label{fig:directions}       
\end{sidewaysfigure}

North-South asymmetries in ecliptic coordinates have been observed in
the WMAP CMB maps using
several local quantities: power spectrum and 2 and 3-point correlation
functions \cite{eriksen_04a, eriksen_05}, Minkowski functionals
\cite{eriksen_04b}, 
local power spectrum \cite{hansen_04a, donoghue_donoghue_05}, local
bispectra \cite{land_magueijo_05b} and local 
curvature \cite{hansen_04b}. Varying the coordinate system, a
maximum asymmetry for the two hemispheres is obtained for a system
whose north pole is lying at $(\theta, \phi) = (80^{\circ},
57^{\circ})$. This direction is close to the north ecliptic pole $(\theta,
\phi) = (60^{\circ}, 96^{\circ})$ (see Fig.~\ref{fig:directions}). The
result of 
all those works basically indicates a significant lack of power in the
north ecliptic hemisphere compared to the south one. The asymmetry
has been also confirmed by \cite{bernui_06, bernui_07} using a pair
angular separation histogram method.

The asymmetry remains stable with respect to variations in the Galaxy
cut and to the frequency band. Also
a similar asymmetry is found in the COBE-DMR map with the axis of
maximum asymmetry close to the one found in the WMAP data. Analyses of
the possible foreground contamination and known 
systematics do not seem to be the cause of such asymmetry \cite{eriksen_04a}. 

More recently, the asymmetry has been found again in the 3-year WMAP
CMB map \cite{eriksen_07}. 

\subsection{Alignment of the low multipoles \label{subsec:alignment}}

\begin{figure}
\centering
\includegraphics[angle=-90, width=1\textwidth]{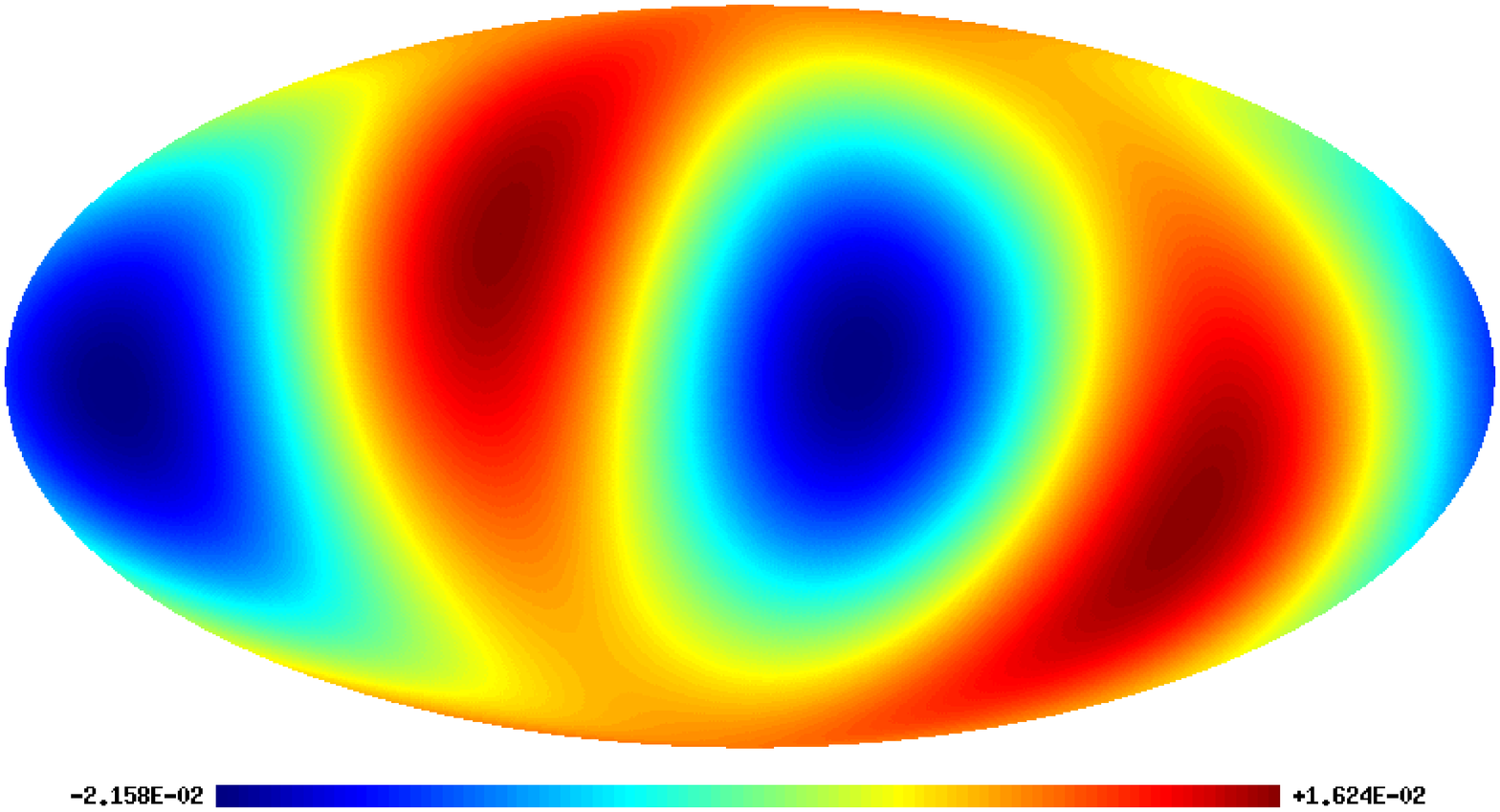}
\includegraphics[angle=-90, width=1\textwidth]{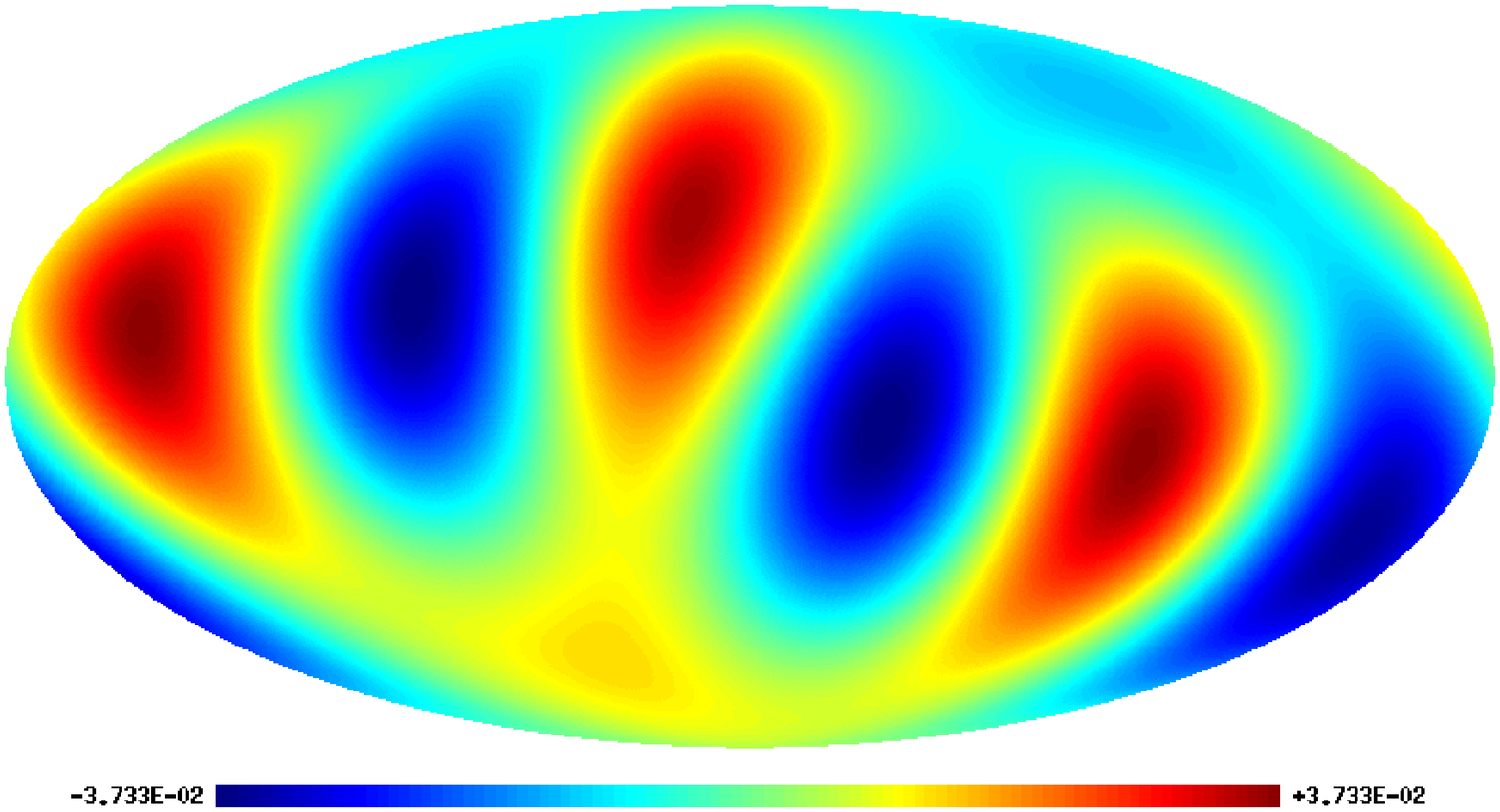}
\caption{Maps of the quadrupole and octopole obtained from the WMAP
  5-years ILC map. The ILC map can be obtained from the LAMBDA web page
  (http://lambda.gsfc.nasa.gov/).} 
\label{fig:ILC-l23}       
\end{figure}

The lowest mulpoles, especially $\ell =2,3$, of the WMAP data have
been found to be anomalously planar and aligned \cite{copi_04,
copi_07, oliveira-costa_04, katz_weeks_04, schwartz_04,
bielewicz_05}. In Fig.~\ref{fig:ILC-l23} the quadrupole and octopole
of the ILC map are shown. Both multipoles present maxima and minima
following a planar shape, whose perpendicular axis points towards
similar direction called axis of evil. The axes of the two multipoles
are separated by $\approx 10^\circ$. The 
probability that the two directions are separated by that angle or
less by chance is $\approx 1.5\%$. Further alignments have also been
claimed 
for higher multipoles $\ell\leq 5$ \cite{land_magueijo_05a} and $\ell
=6,7$ \cite{freeman_06}. The northern end of the alignment points
towards $(\theta, \phi) = (30^{\circ}, 260^{\circ})$, a direction 
close to the CMB dipole one whose northern end is at $(\theta,
\phi) = (42^{\circ}, 264^{\circ})$ (see Fig.~\ref{fig:directions}).

A problem which appears when trying to estimate the low
multipole components is that they are very much affected by the
mask. Varying the mask produces significant changes in their amplitude
estimates, especially for the quadrupole, implying consequently
uncertainties 
in the determination of their axes. Detailed analyses of this effect
tend to weaken the significance of the detection
\cite{oliveira-costa_06, land_magueijo_07}.       

\subsection{The cold spot \label{subsec:cold_spot}}

\begin{figure}
\centering
\includegraphics[angle=-90, width=1\textwidth]{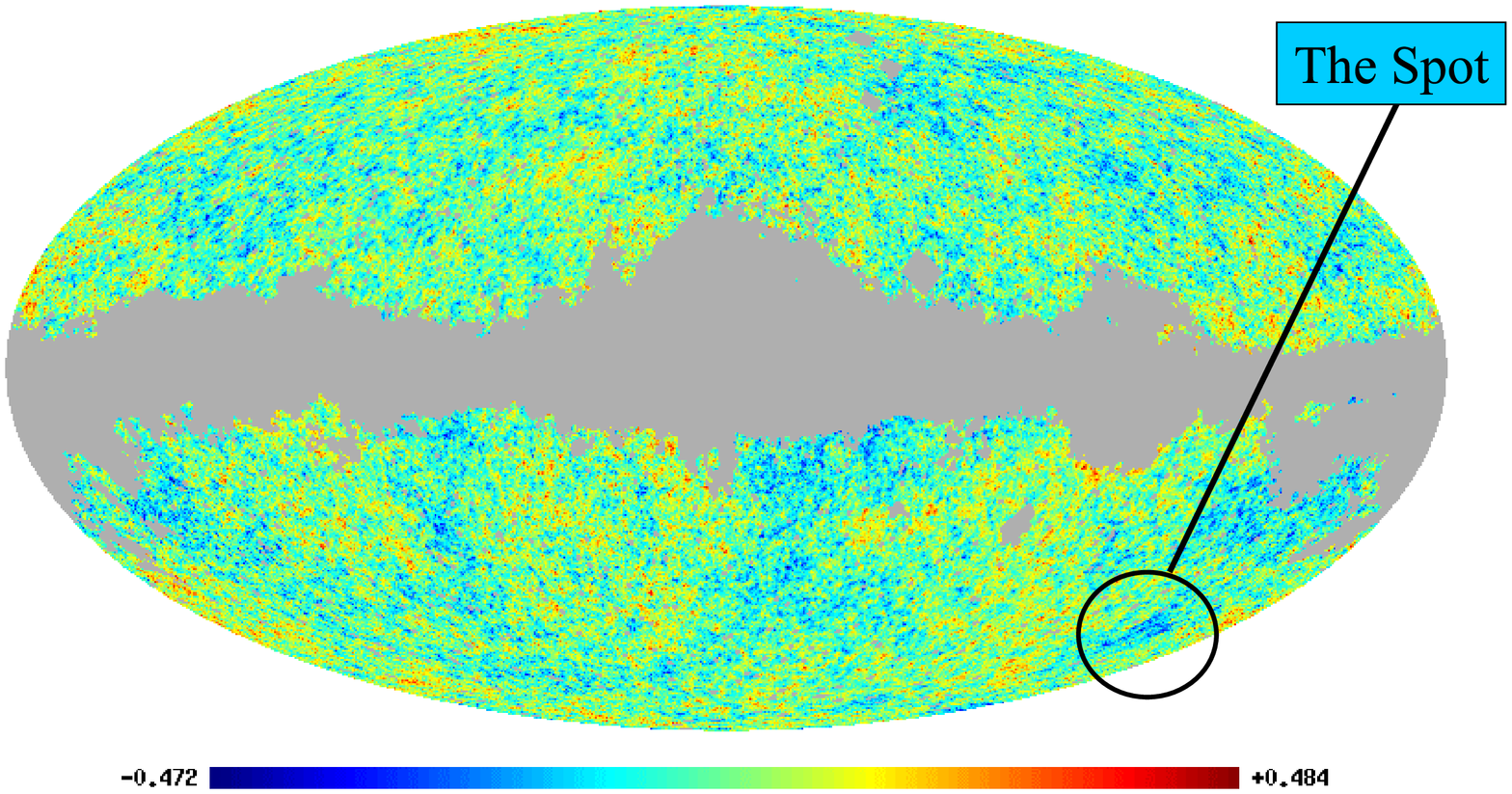}
\caption{The cold spot in the 5-years WMAP combined map. This map is a
  noise-weighted combination of the V and W maps given in  
  \cite{hinshaw_08} where the pixels contaminated by Galactic or
  extragalactic foregrounds have been masked with the KQ75 mask. The
  WMAP maps and masks can be found in the LAMBDA web page.}
\label{fig:cold-spot}       
\end{figure}

A large and prominent cold spot has been observed in the WMAP data
which is hard 
to explain within the standard inflationary scenario. It was detected
with the SMHW as defined in \cite{martinez-gonzalez_02}. The
first evidence came from the kurtosis of the wavelet coefficients of
the first-year data 
which showed an excess with respect to the Gaussian hypothesis at a
wavelet scale of $\approx 4^\circ$ (corresponding to a structure size
of $\approx 10^\circ$) \cite{vielva_04}. A very cold
spot at Galactic coordinates $(\theta, \phi) = (-57^{\circ},
209^{\circ})$ was identified as the possible source of the excess (see
Fig.~\ref{fig:cold-spot}). An analysis
of the area of the spots at different thresholds in the SMHW
coefficient map at around $4^\circ$ proved that indeed the cold spot
had a very large area and was the source of the excess of the kurtosis
\cite{cruz_05}.

The cold spot has been also shown to deviate from Gaussianity using
the Max and Higher Criticism estimators \cite{cruz_05, cayon_05}. A
study of the morphology of the spot with the elliptical 
MHW on the sphere has found an almost circular shape
\cite{cruz_06}. In the same paper the possible foreground contribution
was considered in detail, concluding that contributions from the SZ effect
and the Galaxy had to be negligible. 

The cold spot has been also identified as the most prominent spot in
the CMB sky using steerable wavelets \cite{vielva_07a}. In that
work two other spots are identified as deviations from the IGRF
three-year WMAP best fitting model. However, the deviation from
Gaussianity seen in the 
kurtosis of the coefficients of that wavelet \cite{wiaux_08}
cannot be assigned exclusively to those three spots. 

\cite{mcewen_05, mcewen_08} have also detected a number of non-Gaussian
features, including the cold spot, using the directional wavelets
elliptical Mexican hat and Morlet. 

All the previous results have been confirmed with the
3-year WMAP CMB map \cite{cruz_07} and are expected to be almost
unaltered for the 5-year data.   

\subsection{Alignment and signed-intensity of local structures
  \label{subsec:local_structures}} 

Some of the previous anomalies imply preferred directions in the sky
that, under the assumption of Gaussianity, represent deviations from
statistical 
isotropy. Here we describe a different violation of statistical
isotropy based on the
alignment of CMB structures. The structures are identified by
convolving the CMB map with the
steerable wavelet formed by the second Gaussian derivative
\cite{wiaux_05}. For each scale and position in the sky the wavelet 
identifies the orientation which maximizes the absolute value of the
wavelet coefficients. Thus this orientation corresponds to the
characteristic orientation of the local feature of the signal. The
wavelet coefficient in that specific orientation defines the so-called
signed-intensity \cite{vielva_07a}. It
should be remarked that these two quantities, local 
orientation and signed-intensity, are computationally feasible because
of the steerability property (see Sec.~\ref{sec:methods}). 

\begin{figure}
\centering
\includegraphics[angle=-90, width=1\textwidth]{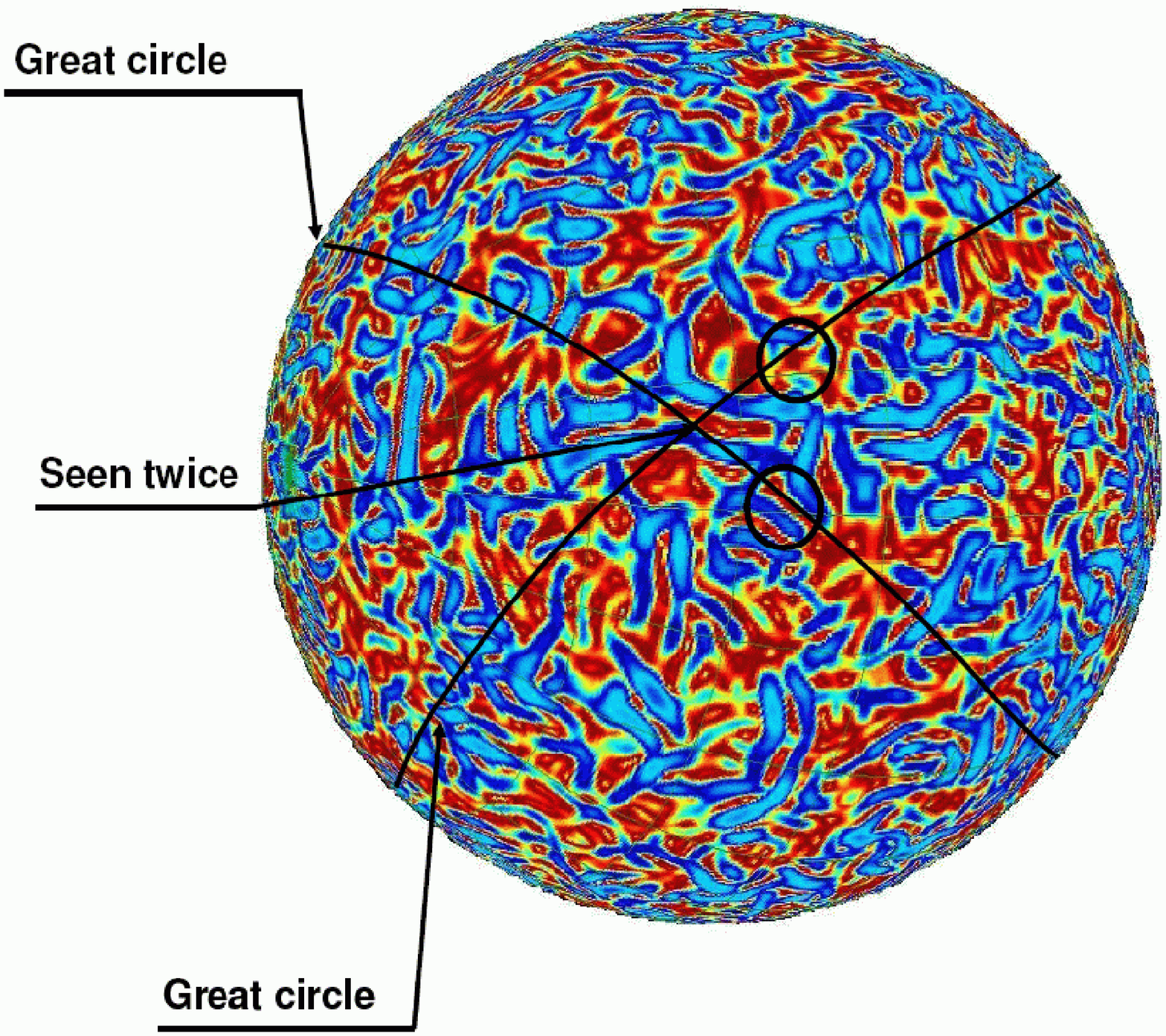}
\caption{A simulated map of the signed-intensity where the great
  circles corresponding to two local features are shown. All the
  pixels crossed by each great circle are considered to be seen by the
  corresponding local feature. The position which is crossed by both
  great circles is thus said to be seen twice. See \cite{vielva_07a}
  for more details.}
\label{fig:local_orientation}       
\end{figure}

Once the local orientation is determined for every position in the sky
at a given scale, we can construct the following isotropy test. First,
the great 
circle passing by that position and tangent to the local orientation
is identified. Every other pixel which is now crossed by that great
circle 
is considered to be seen by the local orientation of the first
position (this procedure is illustrated in
Fig.~\ref{fig:local_orientation}), with a weight naturally given by
the absolute value of the 
signed-intensity of the first position. Second, we can assign to each
pixel in the sky the total weight given by the sum of the weights of
all the pixels which see that pixel. This new total weight signal is
even under parity and thus its analysis can be restricted to one
hemisphere of reference. The highest total weights represent the
positions 
towards which the CMB features are predominantly directed while the
lowest ones represent the positions predominantly avoided by the CMB
features. Of course, for an all-sky IGRF all the pixels should have
the same total weight on average.  

\begin{figure}
\centering
\includegraphics[angle=-90, width=1\textwidth]{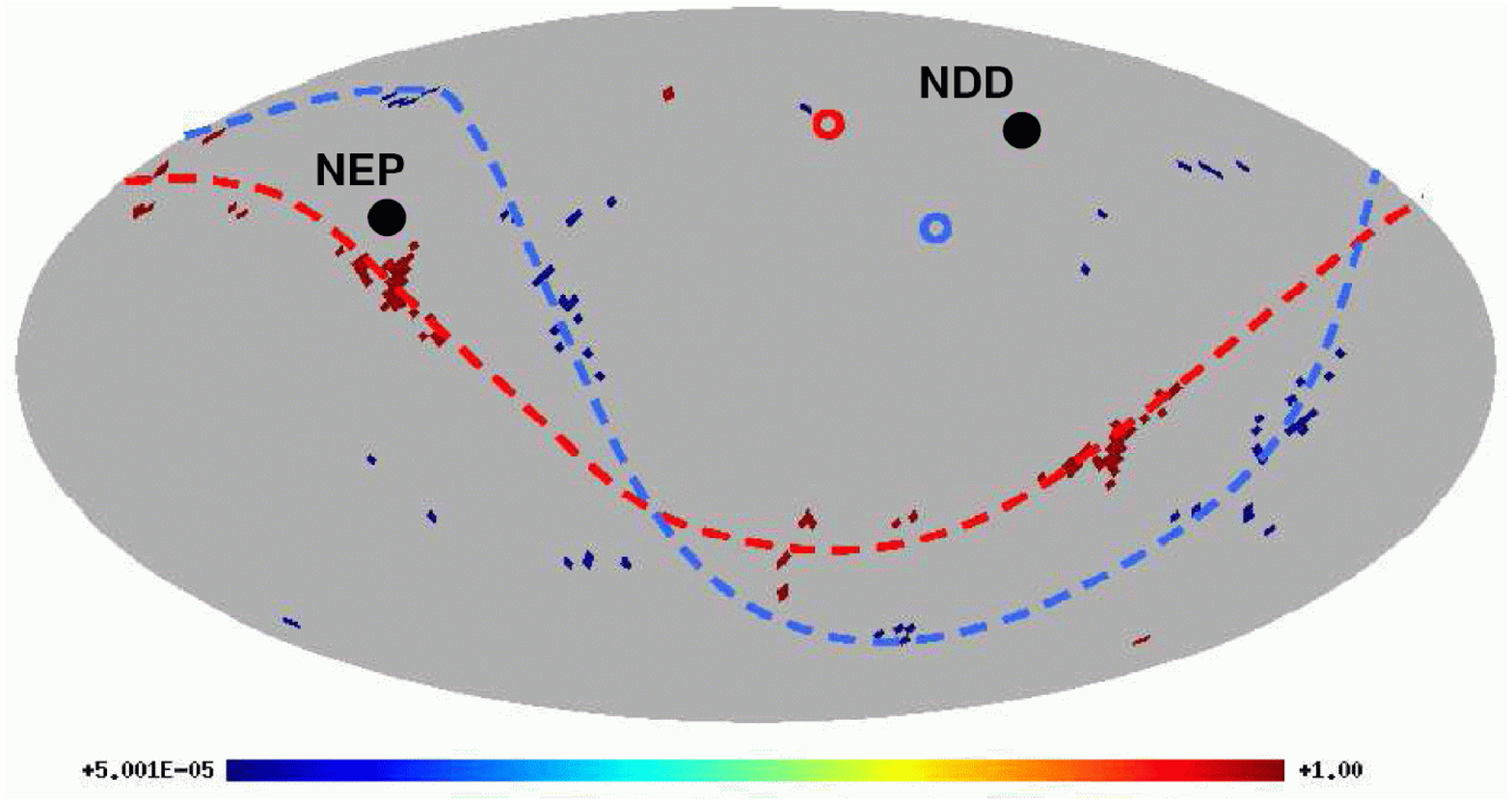}
\caption{The highest (lowest) total weights above (below)
  $3\sigma$ are plotted in red (blue). A cluster of positive
  total weights can be seen close to the NEP. The red (blue) dashed
  line is the best fit plane to the highest (lowest) total
  weights. The normal axes defined by both planes point towards
  directions close to the northern direction of the dipole NDD. From
  \cite{vielva_07a}.} 
\label{fig:alignment}       
\end{figure}

This isotropy test based on the alignment of local orientations has
been applied to the first \cite{wiaux_06} and three-year
\cite{vielva_07a} WMAP data. In both cases a significant violation of 
isotropy was found at a scale around $10^\circ$. The highest total
weights (above $3\sigma$) define an axis located very close to the
ecliptic one.  
The highest and the lowest total weights define two planes whose
normal axes are close to the CMB dipole one (Fig~\ref{fig:alignment}).

\begin{figure}
\centering
\includegraphics[angle=-90, width=1\textwidth]{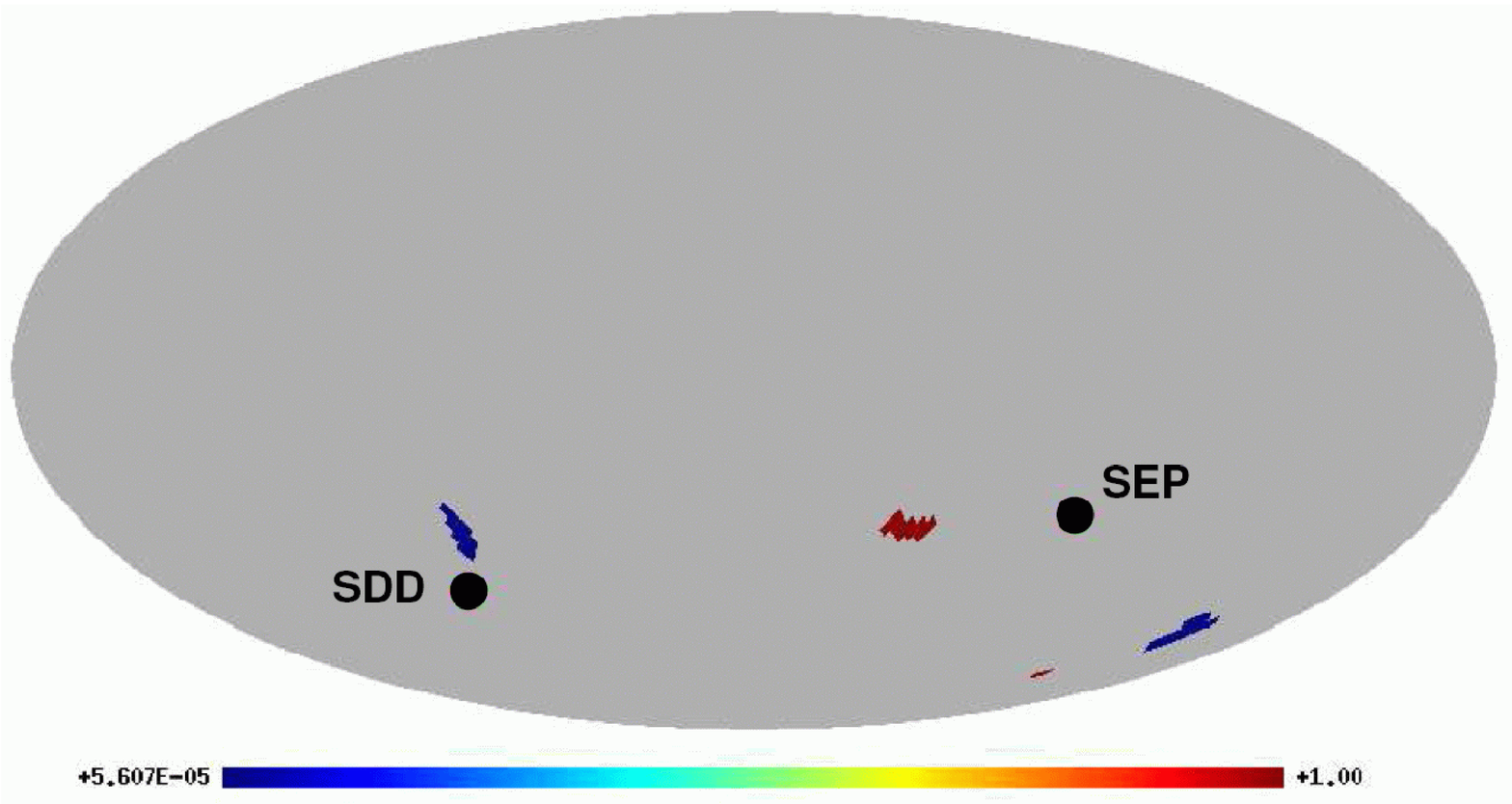}
\caption{Map of signed-intensities with probabiliy in the $3\sigma$
  tails. In blue are the pixels with negative values and in red the
  positive ones. The signed-intensities are grouped around three
  clusters, one formed by the negative values and other two by the
  positive ones. Also shown are the positions of the South ecliptic pole
  (SEP) and the southern direction of the CMB dipole (SDD). From
  \cite{vielva_07a}.}
\label{fig:signed-intensity}       
\end{figure}

Besides the alignment of the local orientations, an analysis of the
signed-intensities show three spots at angular scales also around
$10^\circ$  containing a total of 39 $1^\circ.8$-pixels 
whose values have a (signal+noise) probability in the
$3\sigma$ tails (see Fig.~\ref{fig:signed-intensity}). This pattern is
similar to the one found with the 
axisymmetric Mexican hat wavelet \cite{vielva_04}. The three spots
are located in the southern galactic hemisphere confirming the
north-south asymmetry. Two of the
them are cold, one being identified with the cold spot already
detected in \cite{vielva_04}. 

The two anomalies which appear at similar scales, with global
significance levels around $1$ per cent, are however quite independent.

\subsection{Low variance \label{subsec:low_variance}}

\begin{figure}
\centering
\includegraphics[width=0.6\textwidth]{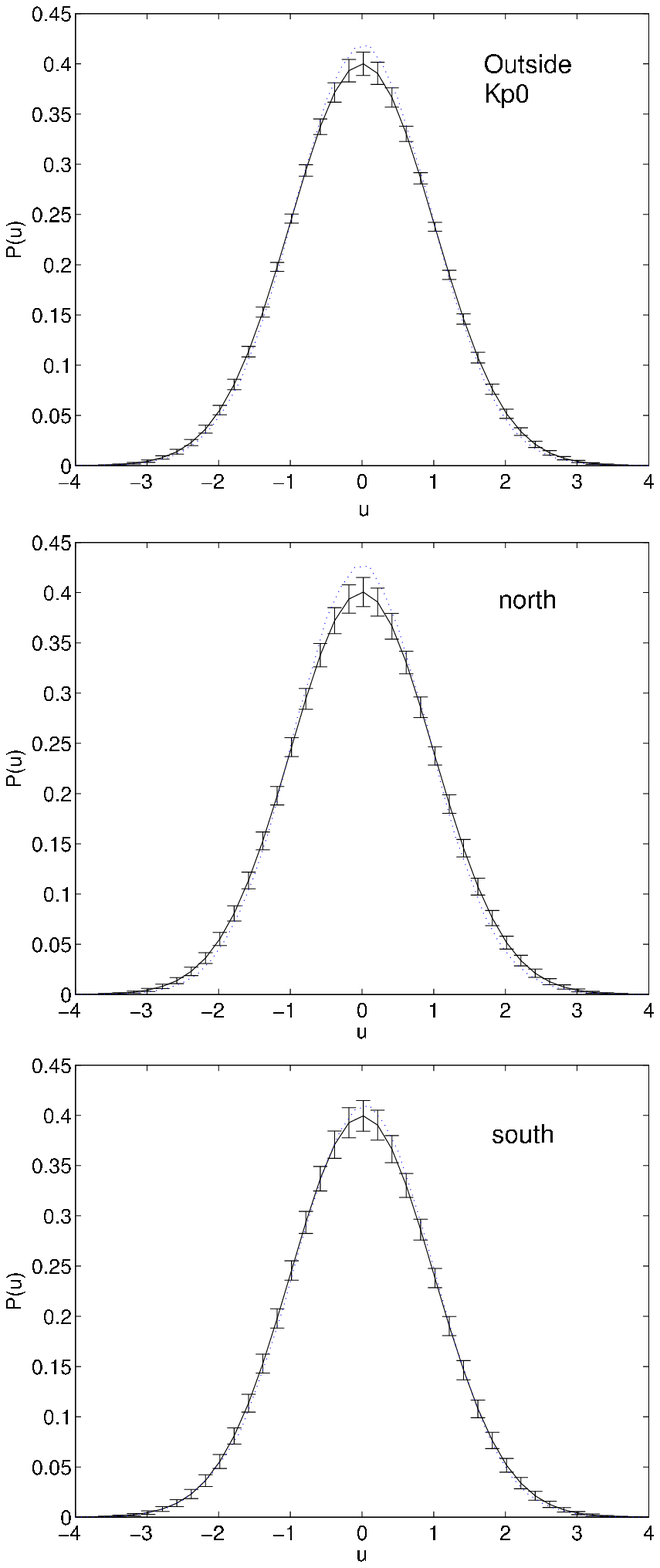}
\caption{The histogram of the normalized WMAP data outside the Kp0
  mask (dotted line) is compared with the average histogram obtained
  from 1000 Gaussian simulations (solid line) in the top panel. The error bars
  represent the 
  dispersion obtained from the simulations. Analogous histograms for
  the northern and southern ecliptic hemispheres are given in the middle and
  bottom panels. See \cite{monteserin_07} for more details.}
\label{fig:pdf-hemispheres}       
\end{figure}

\begin{figure}
\centering
\includegraphics[width=0.7\textwidth]{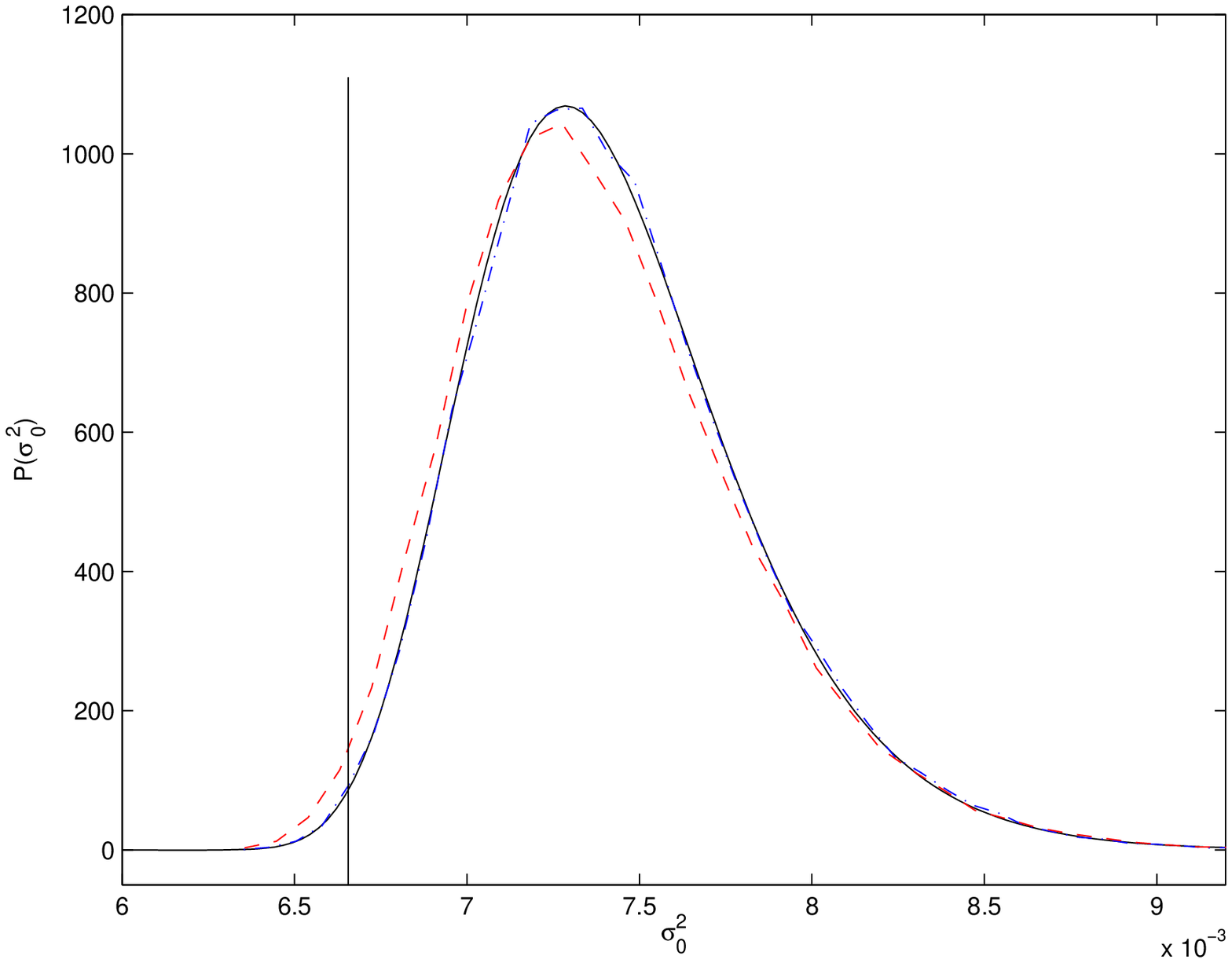}
\caption{Theoretical pdf of the CMB variance (calculated following
  \cite{cayon_91}, solid line) compared
  to the averaged pdf obtained from 60000 Gaussian simulations of the
  WMAP best-fit model over the whole 
  sky (blue dot-dashed line) and using only the pixels outside the
  Kp0 mask (red dashed line). The solid vertical line indicates the
  value obtained from the 3-year WMAP data. See \cite{monteserin_07} for more
  details.} 
\label{fig:variance-pdf}       
\end{figure}

Very recently \cite{monteserin_07} the 1-pdf
of the WMAP three-year data has been analysed finding an anomalous low
value of the 
variance as compared to the one expected from the WMAP best-fit
cosmological model. The result is even more prominent if only the
north ecliptic hemisphere is considered (see
Fig.~\ref{fig:pdf-hemispheres}), in 
agreement with the lack of power found in that 
hemisphere by previous works (see e.g. \cite{eriksen_04a}). The
variance of the CMB signal is obtained by 
fitting the normalized temperature distribution to a Gaussian of zero
mean and unit variance. The significance of the result is around $1\%$
(see Fig.~\ref{fig:variance-pdf}). 

In order to find a possible origin for this anomaly the behaviour of
single radiometer and single year data as well as the effect of
residual foregrounds and 1/f noise have been studied. None of these
possibilities can explain the low value of the variance.

Since the largest contribution to the variance comes from the lower
multipoles, it is interesting to see if the low quadrupole measured by
COBE and 
WMAP is the cause of the anomalously low variance. Performing the same
analysis after subtracting the best-fit quadrupole outside the Kp0
mask the significant of the result is slightly reduced although the
variance is still anomalously low.

Beyond the inconsistency found between the best-fit model and the
measured 
variance, one could ask if the latter is consistent with the actual
measured power spectrum of the WMAP data. The analysis performed by
the same authors show that a strong discrepancy is indeed found. This
last result suggests a possible deviation of the CMB data form the IGRF.

\subsection{Cosmological consequences
  \label{subsec:cosmological_consequences}} 

Given that neither foreground contamination nor known systematics seem
to be causing most of the previous anomalies, there has been a number of 
attempts to explain the cause of the WMAP anomalies by an intrinsic
origin. Among them we mention the Rees-Sciama effect produced by large
voids \cite{tomita_05, inoue_silk_07, rudnick_07}, inhomogeneous
\cite{adler_06, land_magueijo_06} or anisotropic universes
\cite{jaffe_05} and cosmic defects \cite{cruz_07}. Although no further 
evidence has been found for those explanations most of them still
remain as plausible. Below we discuss in more detail two
interesting possibilities: the Bianchi model and the cosmic texture.
     
\subsubsection{Bianchi model \label{subsubsec:bianchi}}

A first interesting attempt to explain the best studied WMAP anomalies
was performed by \cite{jaffe_05, jaffe_06a} who suggested that such
features might be produced by an anisotropic universe, the Bianchi
VII$_h$ model. This type of model has a global anisotropic expansion
and vorticity that produces geodesic focusing and a spiral pattern in
the CMB 
anisotropy at large angular scales. By fitting the free parameters of
that model to the 
WMAP map the large scale CMB anisotropies produced by its
non-standard geometry are
determined. After subtracting that pattern from the WMAP data, the
best studied WMAP anomalies, namely the low-multipole alignments, the
north-south asymmetry and the cold spot, were significantly reduced. 

This result seemed to suggest that the universe was not well represented
by the homogeneous and isotropic FLRW model and that a perturbation in
the form of a Bianchi VII$_h$ was a better representation. However, a
more detailed examination of the best-fitted parameters of the Bianchi
model when a dark energy was included showed values of the dark energy
and matter energy density far from the ones 
measured by many current cosmological tests \cite{jaffe_06b,
bridges_07a}. Considering other Bianchi models with vorticity
and shear, like the Bianchi IX with a closed geometry, do not help
since they do not exhibit geodesic focusing or the spiral pattern.   

Very recently, \cite{bridges_07b} computed the Bayesian evidence of
the Bianchi template when the cold
spot was not included in the analysis (see below for an alternative
interpretation of the cold spot). The result was that the evidence was
now significantly reduced, reinforcing the idea that the cold spot was
likely to be driving any Bianchi VII$_h$ detection.

\subsubsection{The cold spot texture \label{subsubsec:texture}}

\begin{figure}
\centering
\includegraphics[angle=-90, width=0.9\textwidth]{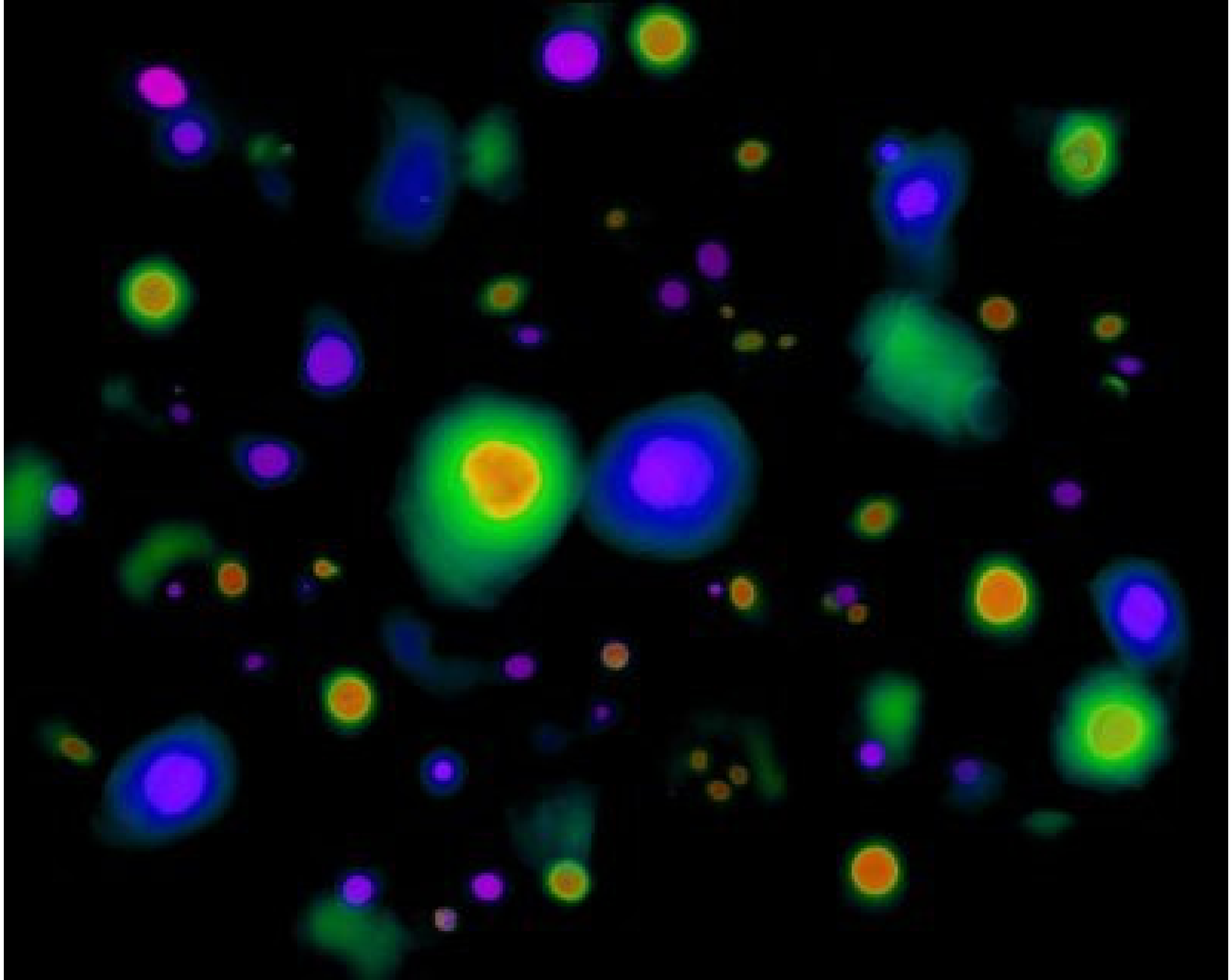}
\caption{Image of the high resolution texture simulation performed by
  N. Turok and V. Travieso and available at
  (http://www.damtp.cam.ac.uk/cosmos/viz/movies/neil.html). The
  distribution of texture spots on the sky is predicted to be scale-invariant.} 
\label{fig:texture_simulations}       
\end{figure}

Recently, \cite{cruz_07} performed a Bayesian evidence
analysis to test the hypothesis that the cold spot is produced by a 
cosmic texture. This is a type of toplogical defect that, as explained
in Sec.~\ref{subsubsec:topological_defects}, in the standard theories
of unification of 
the fundamental forces of nature, are expected to appear in the early
universe. The motivation for considering a texture as the origin of
the cold 
spot is its typical spherical anisotropy pattern left in the CMB and
the relatively small number of spots expected at scales of several
degrees (see Fig.~\ref{fig:texture_simulations}).

The hypothesis test considered consisted in the comparison of the
following two hypotheses: the null hypothesis $H_0$ for which the cold
spot 
is just a rare fluctuation of the IGRF predicted by the standard
inflationary model, and the alternative hypothesis, $H_1$ for which on
top of 
the inflationary Gaussian CMB fluctuations there are non-Guassian ones
as produced by the texture model. The test is applied to a circular
area of $40^\circ$ diameter centered on the cold spot position. 
The posterior probability ratio of the two hypotheses is:
\begin{equation}
\rho=\frac{Pr(H_1|{\bf D})}{Pr(H_0|{\bf
    D})}=\frac{E_1}{E_0}\frac{Pr(H_1)}{Pr(H_0)} 
\ \ ,   
\end{equation}
where $\bf D$ is the data vector, and $E_i$ the evidence which is the average
likelihood $L$ with respect to 
the priors $\Pi(\Theta_i)$ in the parameters $\Theta_i$ of the
hypothesis $H_i$, 
\begin{equation}
E_i=Pr({\bf D}|H_i)=\int L_i(\Theta_i|H_i) \Pi(\Theta_i) d\Theta_i \ \ ,
\end{equation}
and $Pr(H_i)$ is the a priori probability of hypothesis $H_i$. The a
priori probability ratio is usually set to unit for lack of
information, but since in our case the analysis is centered in the
cold spot (an a posteriori selected pixel) it should be given by the
sky fraction $f_s$ covered by textures. Given that a scale-invariant
distribution of spots is expected and considering only textures
above $1^\circ$ (photon difusion would smear out textures smaller than
that) $f_s=0.017$. The likelihood funtion is simply
$L\propto(-\chi^2/2)$ where $\chi^2=({\bf D}-{\bf T})' {\bf N}^{-1} ({\bf
  D}-{\bf T})$ and ${\bf N}$ is the
CMB+noise covariance matrix. The anisotropy pattern $T$ produced by a
texture can be approximated by an analytical spherical profile
\cite{vilenkin_shellard_00} with only two free parameters: the
amplitude which is related to the fundamental symmetry-breaking energy
scale, and the angular scale which is related to the redshift when the
texture unwinds.

\begin{figure}
\centering
\includegraphics[height=6.8in, width=2.3in]{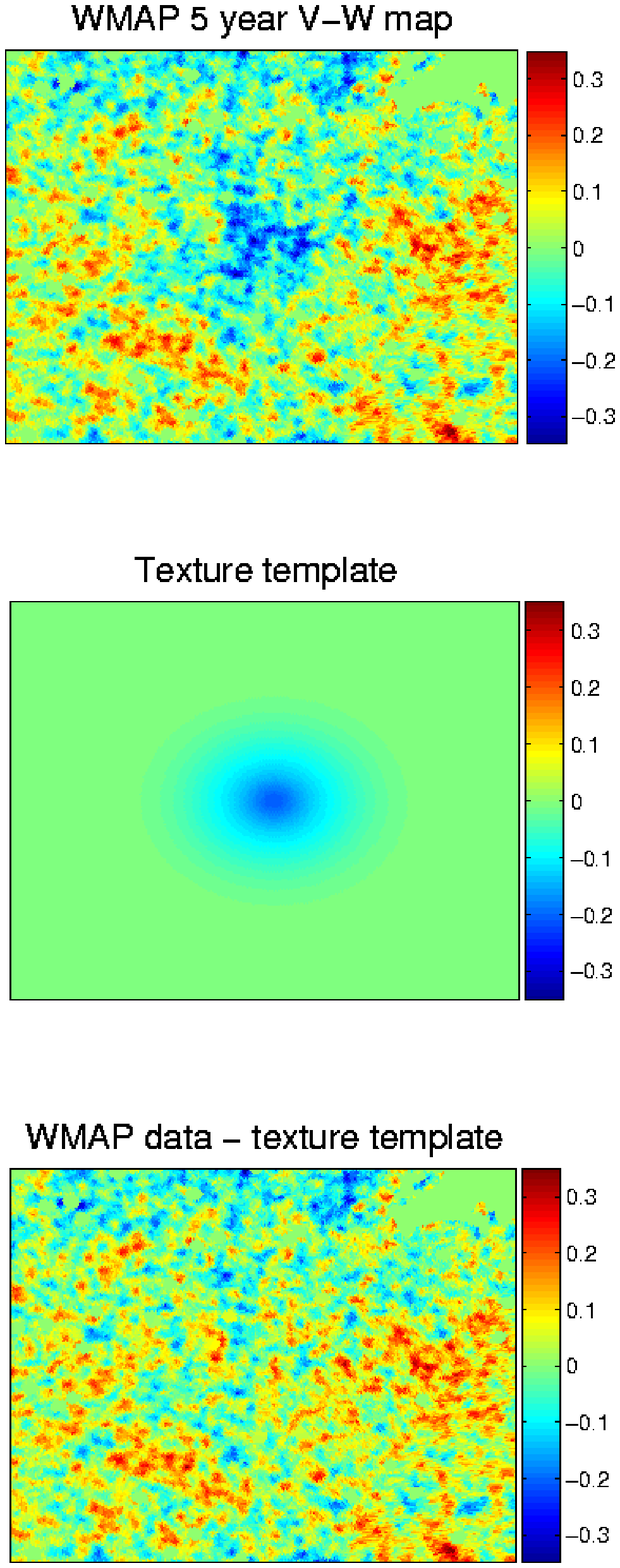}
\caption{A $43^\circ\times 43^\circ$ patch centered at Galactic
  coordinates ($-57^\circ , 209^\circ$) and obtained from the 5-year
  WMAP combined map given in Fig.~\ref{fig:cold-spot}, is
  shown in the top panel. The best fit texture template is in the
  middle and the result of subtracting it from the WMAP map in the
  bottom. The units shown in the colorbars are $mK$. See
  \cite{cruz_07} for more details.}
\label{fig:cold_spot_texture}       
\end{figure}

The result of the analysis is a probability ratio $\rho=2.5$ for the
three-year WMAP data, favouring
the texture hypothesis (see Fig~\ref{fig:cold_spot_texture} for the
resulting best-fit 
texture template). This result is slightly increased ($\rho=2.7$)
for the five-year data due to the reduction in the noise
amplitude. The texture interpretation helps to alleviate the excess
found in the kurtosis, being at the same time compatible with the
observed 
abundance, shape, size and amplitude of the spot. In particular, the
symmetry-breaking scale inferred from this analysis, $\phi_0\approx
8.7\times 10^{15}$GeV, is compatible with the upper limit obtained from
the CMB power spectrum analysis \cite{bevis_04}.

More recently, \cite{cruz_08} have extended the Bayesian evidence
analysis to models based on the Rees-Sciama effect produced by voids
or on the SZ effect. The result is that, contrary to the texture
model, no positive evidence is found for those models.    
  
\section{Concluding remarks and future perspectives\label{sec:conclusions}}

The main conclusion derived from the large amount of analyses,
involving a variety of methods,
performed to test the Gaussianity of the CMB anisotropies, specially
those measured by WMAP, is that the standard IGRF prediction is a good
representation of their properties as a first approach. Furthermore, a
number of significant deviations from the ideal IGRF has also been
reported whose origin and interpretation is still under debate. Some
of them might have to do with foreground residuals or unknown
systematics while others could be a hint of new physics with profound
implications for our understanding of the universe. Probably, to answer
those questions we will have to wait for new data coming from the
advanced experiments being built and expected to be operative in the
next years. 
 
Future experiments will shed light on the open questions remaining
from the up-to-date analyses of the CMB data, specially on the WMAP
anomalies discussed in Sec.~\ref{sec:WMAP_anomalies}. In particular, the
Planck mission is expected to provide all-sky, high quality,
multifrequency maps in the frequency range $30-900$GHz. The wider
frequency range and the higher sensitivity and resolution will
allow an improvement in the quality of the resulting CMB map. As a
consequence, an improvement in the control of the foreground  emission
is expected as well as a reduction in the sky area required to be
masked. In addition to the temperature, improvements are also expected in
the polarization maps which will be provided by Planck, meaning an
important complement for probing 
the nature of the anomalies as well as for testing the different physical
interpretations proposed for them. Missions for measuring
polarization at the highest sensitivity allowed by present technology,
and with the main aim of probing the existence of the gravitational wave
background, have been recently proposed to both agencies, ESA and
NASA. As for the temperature, the linear polarization expected to be
produced as a consequence of the standard inflationary period of the
universe also posseses properties very close to those of the IGRF studied in
Sec.~\ref{sec:IGRF}. Therefore, extensions of the methods already
discussed in Sec~\ref{sec:methods} for temperature are also expected to be    
applied to test the Gausianity of the future polarization maps.

\vskip 1.true cm

{\noindent{\bf Acknowledgements}}\\
I thank R.B. Barreiro and P. Vielva for useful comments on the
manuscript and M. Cruz, A. Curto and C. Monteser\'\i n for helping me
with some figures.  I
acknowledge financial support from the Spanish MEC project
ESP2007-68058-C03-02.  I also acknowledge the use of LAMBDA, support
for which is provided by the NASA Office of Space Science. The work
has also used the software package HEALPix
(http://www.eso.org/science/healpix) developed by K.M. Gorski,
E.F. Hivon, B.D. Wandelt, J. Banday, F.K. Hansen and M. Barthelmann
\cite{gorski_05}.

\end{document}